\documentclass[a4paper,11pt]{article}
\pdfoutput=1

\newcommand{\Msun}{\ensuremath{M_\odot}}
\newcommand{\Mdot}{\ensuremath{\dot{M}}}

\newcommand{\MBH}
{\ensuremath{M_\mathrm{BH}}}
\newcommand{\MH}
{\ensuremath{M_\mathrm{h}}}
\newcommand{\fPBH}
{\ensuremath{f_\mathrm{PBH}}}
\newcommand{\veff}{\ensuremath{v_\mathrm{eff}}}

\newcommand{\rB}{\ensuremath{r_\mathrm{B}}}

\newcommand{\rBeff}{\ensuremath{r_\mathrm{B}^\mathrm{eff}}}

\newcommand{\rh}{\ensuremath{r_\mathrm{h}}}
\newcommand{\cs}{\ensuremath{c_\mathrm{s}}}
\newcommand{\csin}{\ensuremath{c_\mathrm{s, in }}}
\newcommand{\vin}{\ensuremath{v_\mathrm{in}}}
\newcommand{\vrel}{\ensuremath{v_\mathrm{rel}}}
\newcommand{\vR}{\ensuremath{v_\mathrm{R}}}
\newcommand{\vD}{\ensuremath{v_\mathrm{D}}}
\newcommand{\rhoin}{\ensuremath{\rho_\mathrm{in }}}

\def\beq{\begin{equation}}
\def\eeq{\end{equation}}
\def\bea{\begin{eqnarray}}
\def\eea{\end{eqnarray}}

\usepackage{jcappub} % for details on the use of the package, please
\usepackage{enumerate}
\usepackage{tikz}
\usepackage{enumitem}
\usepackage{booktabs}
\usepackage{xspace}
\usepackage{subcaption} % for subfigure
\graphicspath{figs/}
\usepackage[justification=justified,singlelinecheck=false,font=small]{caption}
\usetikzlibrary{arrows,positioning}

\usepackage{aas_macros} % For journal abbreviations

\begin{document}

\title{Feedback in the dark: a critical examination of CMB bounds on primordial black holes}

\author[a]{Dominic Agius,}
\author[b]{Rouven Essig,}
\author[c]{Daniele Gaggero,}
\author[d]{\\Francesca Scarcella,}
\author[b]{Gregory Suczewski,}
\author[e]{Mauro Valli}

\affiliation[a]{Instituto de Fisica Corpuscular (IFIC), Universidad de Valencia-CSIC, 46071, Valencia, Spain}
\affiliation[b]{C.N. Yang Institute for Theoretical Physics, Stony Brook University, NY 11794, USA}
\affiliation[c]{INFN Sezione di Pisa, Polo Fibonacci, Largo B. Pontecorvo 3, 56127 Pisa, Italy}
\affiliation[d]{Laboratoire Univers et Particules de Montpellier (LUPM), Université de Montpellier \& CNRS, 34095 Montpellier, France}
\affiliation[e]{INFN Sezione di Roma, Piazzale Aldo Moro 2, I-00185 Rome, Italy}

\emailAdd{dominic.agius@ific.uv.es}
\emailAdd{gregory.suczewski@stonybrook.edu}
\emailAdd{francesca.scarcella@umontpellier.fr}

\abstract{
If present in the early universe, primordial black holes (PBHs) will accrete matter and emit high-energy photons, altering the statistical properties of the {Cosmic Microwave Background (CMB)}. This mechanism has been used to constrain the fraction of dark matter that is in the form of PBHs to be much smaller than unity for PBH masses well above one solar mass. Moreover,  the presence of dense dark matter mini-halos around the PBHs has been used to set even more stringent constraints, as these would boost the accretion rates.  
In this work, we critically revisit CMB constraints on PBHs taking into account the role of the local ionization of the gas around them. We discuss how the local increase in temperature around PBHs can prevent the dark matter mini-halos from strongly enhancing the accretion process, in some cases significantly weakening previously derived CMB constraints. We explore in detail the key ingredients of the CMB bound and derive a conservative limit on the cosmological abundance of massive PBHs.
}   

\keywords{primordial black holes, cosmic microwave background, dark matter, accretion}

\maketitle

\section{Introduction}\label{sec:introduction}

The cosmic dark ages, situated between recombination and reionization, were  characterized by a very low ionization fraction. During this period, the energy density of the universe was dominated by adiabatically cooling dark matter (DM) and baryons, with the latter mostly in the form of neutral hydrogen. No astrophysical processes involving copious emission of thermal or non-thermal radiation are expected to have been 
present in this epoch, which makes it {a powerful probe of new physics that produces such radiation}.

Any extra energy injection, such as particle DM annihilation or decay, can significantly heat, excite, and ionize the baryonic gas. These alterations of the baryonic medium can modify the optical depth and the visibility function of the relic photons that decoupled from the baryons shortly after recombination, ultimately affecting the anisotropies of the cosmic microwave background (CMB). This powerful way to constrain new physics was first discussed in a series of pioneering papers at the beginning of the century~\cite{Peebles:2000pn,Pierpaoli_2004,Chen:2003gz,Padmanabhan_2005,Zhang_2006,Zhang_2007}. 

Primordial black holes (PBHs), which may constitute a component of DM, can be probed in a similar way. If massive enough, PBHs will efficiently accrete the surrounding medium, heating the baryons and emitting a broad spectrum of ionizing photons that can alter the CMB power spectrum. This idea was first discussed in~\cite{1981MNRAS.194..639C} and later developed, for instance in~\cite{Ricotti:2007jk,Chen:2016pud,Horowitz:2016lib,AliHaimoudKamionkowski2017,poulin_cmb_2017,Serpico:2020ehh, Facchinetti2022}, setting stringent upper limits on the cosmological abundance of PBHs of mass above $\mathcal{O}(10) \, M_{\odot} $.

However, these works relied on the classic Bondi-Hoyle-Lyttleton (BHL) model~\cite{Hoyle1939, 1952MNRAS.112..195B}  to quantify the rate of gas accretion onto PBHs, a central ingredient in the modelling of the energy injection into the medium. 

The BHL model exists as an interpolation between two regimes: the first being the ballistic regime, considering accretion onto a point mass moving with a constant velocity through a gas of constant density and considering only gravitational effects \cite{Hoyle1939}, and the second being the accretion onto a spherically symmetric body at rest including the effects of pressure and gravitation \cite{bondi_spherically_1952}. It is emphasized that this interpolation formula should only be considered as an order-of-magnitude estimate of the accretion rate \cite{poulin_cmb_2017}. Moreover, given its simplified assumptions, the BHL model is unable to account for the role of radiation feedback, which is expected to play a crucial role in the physics of accretion. 
It is therefore compelling to move beyond the BHL formalism and to take into account the fluid and radiation effects that may significantly alter the behavior of the accretion rate as a function of the PBH speed and the properties of the medium. 
{To this end}, the Park-Ricotti model~\cite{park_accretion_2011,park_accretion_2012,park_accretion_2013} offers a more realistic prescription for the study of accretion effects in PBHs.
{The authors of~\cite{park_accretion_2011,park_accretion_2012,park_accretion_2013}} performed hydrodynamic simulations of accretion onto a compact object, {taking into account the effects of photo-heating and photo-ionization of the environment}. 
These simulations captured the formation of an ionization front, in some cases preceded by a shock wave, and {found} a strong suppression of the accretion rate for low values of the {PBH} velocity. This behavior, more complex than that predicted by the classic BHL model, is well reproduced by the analytical model proposed by the authors (henceforth {referred to as} the PR model).
The PR model can be regarded as the current state-of-the-art description of the accretion process onto isolated compact objects.

Given the expected significant departure of the accretion rate from the standard treatment, it is compelling to revisit the CMB bounds on PBHs taking into account the role of radiation feedback. This was recently done in~\cite{Facchinetti2022}, where this constraint was re-derived using the PR model. The authors showed that the bound could be relaxed by combining the PR model with conservative assumptions about the energy emission mechanism. 

Irrespective of the accretion model used, PBHs are not allowed to constitute all of the DM (in the mass range where the constraint applies). Therefore, the effects of the remaining DM, whatever its nature, must be accounted for. Dense DM structures with steep density profiles are expected to build up around PBHs, starting soon after their formation. It has been argued~\cite{{Mack:2006gz,Serpico:2020ehh}} that the gravitational potential produced by these mini-halos can significantly boost the accretion rate. CMB constraints {on the abundance of PBHs} have then been re-derived {taking} this into account~\cite{Serpico:2020ehh}, resulting in a significant strengthening of the constraint, by up to three orders of magnitude. However, Ref.~\cite{Serpico:2020ehh} relied on the BHL model to describe the accretion rate, ignoring the effect of radiation feedback.
It is natural to inquire whether this strong boost on the accretion rate from 
DM mini-halos still occurs in the presence of a hot, ionized region around the PBH, which is known to dramatically affect the accretion process.

Motivated by these considerations, we address in the present paper for the first time the {\it combined impact} of  two key aspects: {\it (i)} the role of the radiation feedback (as implemented in the PR model), and {\it(ii)} the impact of DM mini-halos on CMB accretion constraints on the PBH abundance.
To do so, we consistently model the role of DM mini-halos within the PR model, and obtain the corrected accretion rate and the corresponding constraints on the fraction of DM in the form of PBHs, $\fPBH$. 
The novel constraint obtained in this way is compared in detail to the ones obtained under simpler assumptions: the BHL model in the absence of DM mini-halos, the BHL model in the presence of DM mini-halos, and the PR model in the absence of DM mini-halos.

Besides the crucial issue of the interplay between the accretion rate modeling and the presence of the DM mini-halos, it is important to recognize that other sources of uncertainty are very relevant in the problem under consideration. 
In particular, we will consider three additional sources of uncertainty: 
\begin{itemize}
    \item The first uncertainty is specific to the PR model, which requires a choice of sound speed within the ionized region. 
\item The second unknown resides in the mechanism that transforms the gravitational energy of the accreted material into heat and ionizing radiation, which eventually alters the  evolution of the ionization fraction of the universe. 
Early works considered simplified emission models, based on the assumption that the accretion flow is spherically symmetric and thermal bremsstrahlung is the dominant radiation process.
However, recent developments have highlighted the crucial role of accretion disks~\cite{poulin_cmb_2017,Serpico:2020ehh}. Accretion disks are ubiquitous in the universe, and plausible arguments suggest that they can form around PBHs as early as the dark ages. The processes {that are} responsible for the transport of energy and angular momentum in the disk and {that} control the non-thermal emission of radiation are {highly complex and lead to a large systematic uncertainty on the CMB bound on PBHs}. 
In particular, the amount of energy transferred to the leptonic part of the accretion flow plays a pivotal role, since the electrons are responsible for most of the emission of the ionizing radiation 
(see, e.g., the discussion in~\cite{Xie_2012}). 
A detailed assessment of the impact of this key ingredient helps determine the robustness of the CMB bound.
\item The third major unknown is the actual deposition of the injected energy into the ambient medium. 
The energy can be deposited via three main channels, ionization, excitation, and heating, all of which are captured by the redshift-dependent energy deposition functions.
Several parameterizations exist in the literature~\cite{Slatyer:2012yq,slatyer_indirect_2016,Finkbeiner_2012}, which we will use to quantify this uncertainty.
\end{itemize}

Given this complicated tapestry of astrophysical effects, we present here a comprehensive discussion of the robustness of the CMB bound on the PBH abundance with respect to these ingredients. We systematically assess how the constraint is affected by all the elements discussed so far, and show the impact on the posterior probability distribution of $\fPBH$ of the aforementioned elements that affect the bound, with special focus on the radiative efficiency and the energy deposition function.

The rest of the paper is structured as follows. 
In \autoref{sec:accretion}, we describe the details of the accretion models considered in this work in the dark ages background cosmology. In \autoref{sec:mini-halos}, we discuss the role of DM mini-halos. In \autoref{sec:PhysicsBound}, we describe the key ingredients needed to calculate the CMB constraints. In \autoref{sec:resultsBounds}, we present the main results of our work, illustrating the interplay between the impact of the radiation feedback and the DM mini-halos.  In \autoref{sec:resultsAstro}, we assess the role of astrophysical uncertainties on the CMB bound. In \autoref{sec:discussion}, we comment about future improvements of our analysis and present our conclusions. 

\section{Accretion models in the cosmological setting}
\label{sec:accretion}
The first step towards the derivation of the cosmological bound is the modelling of the rate of accretion of baryonic gas onto PBHs during the dark ages. In this epoch (the most relevant redshift range is $z \sim 100-1000$) structures have not yet formed. As was done in previous works~\cite{poulin_cmb_2017,Serpico:2020ehh}, we assume that density variations in the accreted gas are negligible and that the gas can be treated as a homogeneous medium, whose temperature and density decrease adiabatically with time according to the cosmological model. We assume that PBHs are distributed smoothly in the universe following the DM distribution. The PBH-gas velocity is then expected to trace the cosmological baryon-DM velocity. We assume that the energy injection can be considered spatially homogeneous. 
This picture is expected to break down at later times, when virialized halos start to form, but is typically considered reliable in the redshift range that is relevant for the CMB bound. We mention in the final discussion possible caveats associated with these assumptions,  related to the early formation of structures in $\Lambda$PBH cosmologies.
Finally, we assume that PBHs are characterized by a monochromatic mass distribution.\footnote{Extended mass functions, as predicted by simple single-field inflationary scenarios, have been shown to typically tighten the CMB bounds on PBH abundance~\cite{Bellomo:2017zsr,K_hnel_2017,Carr_2017,poulin_cmb_2017}. We will comment on this further in \autoref{sec:discussion}.}

In this section, we first review the background cosmology that describes the evolution of the baryonic medium and the DM component with redshift, based on which we model the accreted gas and the PBHs. We then review the classic BHL model before summarizing the main aspects of the PR model, on which we rely to account for the role of radiative feedback. 

\subsection{The background cosmology}
\label{sec:physicalScenario}
The background cosmological density of baryons $\rho_{\mathrm{b}}$ scales as 
\begin{equation}
\label{eq:rho}
\rho_\mathrm{b} = \rho_{\rm b,rec} \frac{\left(1 + z  \right)^3}{\left(1 + z_{\rm rec}  \right)^3} \ ,
\end{equation}
where the baryon density at recombination is $\rho_{\rm {b},rec} \simeq 200 \,\, {\rm cm^{-3}}$ {and the redshift of recombination is $z_{\rm rec} \sim 1100$}. 
The sound speed of the baryons $c_{\mathrm{s}}$ depends on the background temperature $T$ and on the ionization fraction $X_\mathrm{e}$, such that
\begin{equation}
\label{eq:soundspeed}
\cs = \sqrt{\frac{\gamma \left(1 + X_\mathrm{e} \right)T}{m_{\mathrm{p}}}} \ \mathrm{km} \  \mathrm{s}^{-1} \, ,
\end{equation}
where $m_\mathrm{p}$ is the proton mass and $\gamma$ is the adiabatic index.
We take the average velocity between PBHs and the background gas $\vrel$ to be equal to the linear cosmological baryon-DM relative speed~\cite{Tseliakhovich2010,Dvorkin2013}, which is given by 
\begin{equation}
\label{eq:vrel}
\sqrt{  \left \langle \vrel^2 \right \rangle } = {\rm min} \left[ 1, \frac{1+z}{1000}  \right] \cdot 30 \, \mathrm{km} \  \mathrm{s}^{-1} \ .
\end{equation}
The PBH velocity distribution is assumed to be described by a Maxwell--Boltzmann distribution centered around this mean value. 

\subsection{The Bondi--Hoyle--Lyttleton model}
\label{sec:accretion:models:BHL}
According to the BHL accretion model~\cite{Hoyle1939,bondi_spherically_1952}, the accretion rate onto an isolated compact object, characterized by its mass $M$ and speed $\vrel$, is given by
\begin{equation}
\label{eq:bondi}
\dot{M}_\mathrm{BHL} = 4 \pi \lambda \frac {(GM)^2 \rho_{\rm b}} {(\vrel^2 + c_{\mathrm s}^2)^{3/2}} \,\,\, ,
\end{equation}
where $\rho_{\rm b}$ and  $c_{\mathrm s}$ are the ambient medium density and sound speed, respectively, and $\lambda $ is a constant suppression factor (see below). 
The BHL accretion rate can also be expressed in terms of an effective cross section
\begin{equation}
\Mdot_\mathrm{BHL}  = 4 \pi \lambda \rho_{\rm b} \dfrac{G^2 M^2}{\veff ^3} = \,  4 \pi \lambda \rho_{\rm b} \, \veff \, \rB^2  \, ,
\end{equation}
where the effective velocity is $\veff = \sqrt {\cs^2 + \vrel^2}$ and $\rB$ is the generalized Bondi radius
\begin{equation} 
\label{eq:bondiradius}
\rB = \dfrac{G M}{\veff ^2} \,.
\end{equation} 

In several works, both in astronomical and cosmological settings, the BHL formula is corrected ad-hoc by a suppression factor $\lambda$.
In the astronomical context, a value of $\lambda \sim 10^{-2}-10^{-3}$ is included to match 
observations of the luminous emission from accreting compact objects, such as the non-observation of a large population of isolated neutron stars~\cite{Perna_2003} or astrophysical BHs~\cite{Fender:2013ei} in the local region, as well as the studies of nearby active galactic nuclei~\cite{Pellegrini_2005}, and of the supermassive BH at the Milky-Way center~\cite{Wang:2013dqq}. This factor aims to phenomenologically capture the impact of a variety of non-gravitational forces (pressure, fluid viscosity, radiation feedback, etc.), which may partially counteract the gravitational pull of the compact object and suppress the accretion rate. 

Similarly, in the cosmological context, the correction to the Bondi formula due to the coupling of the gas to the CMB photon fluid (accounting for Hubble expansion and DM over-densities) has been computed and explicitly expressed as a function of several thermodynamic quantities that characterize the ambient medium~\cite{Ricotti:2007jk,RicottiOstrikerMack2008ApJ,AliHaimoudKamionkowski2017}. These papers assume spherical symmetry for the accretion process and bremsstrahlung (free-free) radiation near the
Schwarzschild radius as the dominant cooling process.

In the following, we use as a benchmark scenario $\lambda=0.01$ in the context of disk accretion.  
{However, for the spherical accretion case, which leads to weaker bounds and can be considered a conservative scenario, we will instead use the results of~\cite{AliHaimoudKamionkowski2017} to set the suppression factor $\lambda$. 
} 

\begin{figure}[t!]
\centering
\includegraphics[width=0.87\columnwidth]{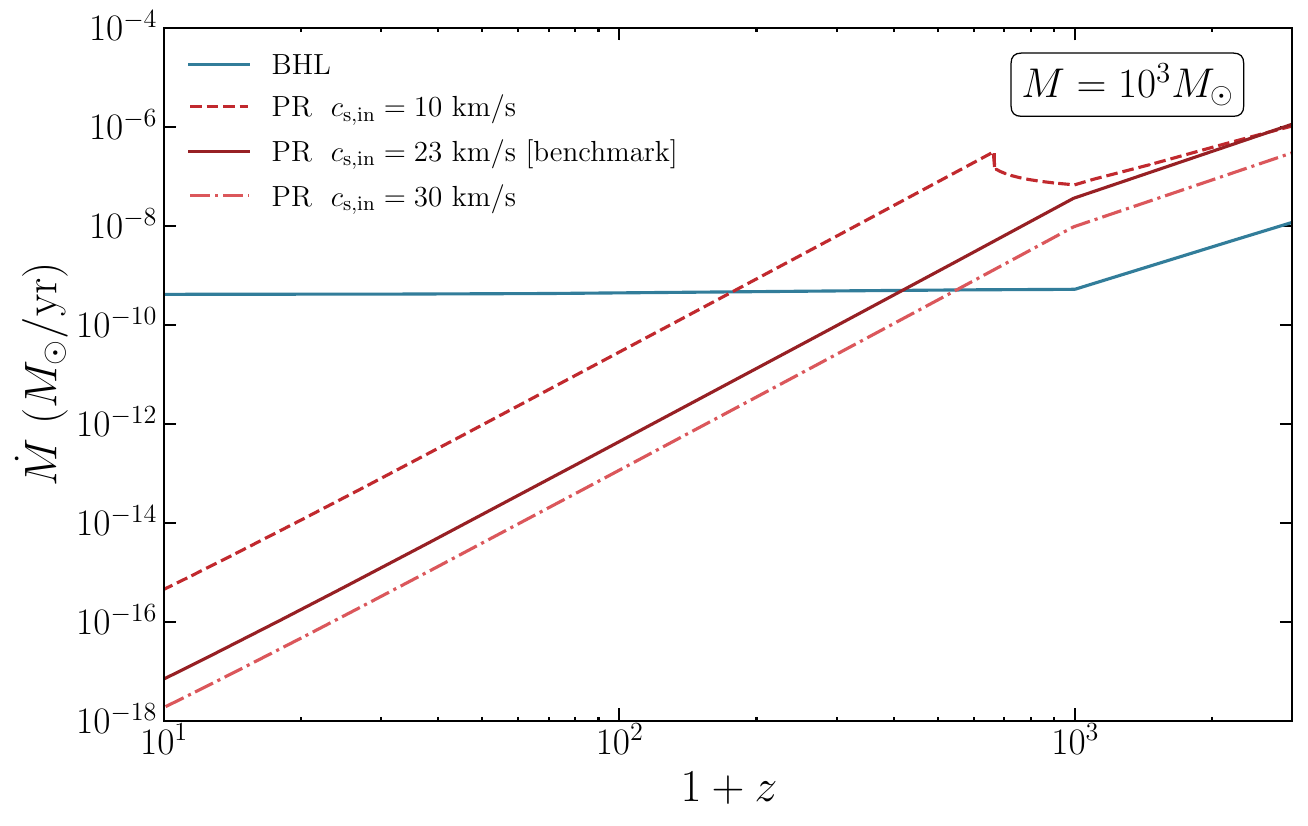}
\caption{Accretion rate as a function of redshift  for the BHL model (blue) and the PR model (red) for a $10^3 \Msun$ PBH, {assuming the setup described in \autoref{sec:physicalScenario}}. For the PR model, three different values are shown for the sound speed within the ionized region $\csin$. The $\Lambda$CDM parameters used for this plot, as well as plots below where a fixed cosmology is implied, are given by: $\left\{ \Omega_\mathrm{b} ,\Omega_\mathrm{c},H_0, \ln10^{10}A_\mathrm{s}, n_\mathrm{s},z_\mathrm{reio}   \right\} = $ $\left\{0.02231,0.1198, 67.90,3.043,0.9674,  7.4 \right\} $. For this plot we set $\fPBH = 10^{-15}$ to remove feedback effects on the background cosmology. }
\label{fig:mdot}
\end{figure}

\subsection{The Park-Ricotti accretion model}
\label{sec:accretion:models:PR}
The impact of \textit{radiative feedback} on the accretion process was studied by Park and Ricotti through hydro-dynamical simulations of accretion of a homogeneous medium into a compact object~\cite{park_accretion_2011,park_accretion_2012,park_accretion_2013}. The simulations showed the formation of an ionization front, in some velocity regimes preceded by a shock wave, and simultaneously a sharp reduction of the accretion rate. The authors were able to reproduce the accretion rates obtained in the simulations via a simple analytical model, the PR model, which we briefly describe here.

The high-energy radiation emitted in the accretion process ionizes the gas surrounding it. Its temperature and density are affected, as well as the velocity with which it flows around the BH. 
This variation in the properties of the gas in turn is expected to affect the accretion rate.
The PR model assumes that the BHL accretion formula, Eq.~\eqref{eq:bondi}, is valid within the ionized region
\begin{equation}
\label{eq:ricotti}
\dot{M}_\mathrm{PR} = 4 \pi \frac {(GM)^2 \rhoin} {(\vin^2 + \csin^2)^{3/2}} \,\,\, ,
\end{equation}
where $\rhoin, \vin, \csin$ are, respectively, the density, {relative} velocity, and sound speed characterizing the ionized gas. The sound speed $\csin$ is set to a constant value, with $\csin \gg \cs$. The density $\rhoin$ and relative speed $\vin$ are obtained from the corresponding quantities $\rho_{\rm b}$ and $v_{\rm rel}$, by imposing the one-dimensional conservation of mass and the equilibrium of forces across the ionization front (we assume an ideal gas and that the process is isothermal \cite{park_accretion_2013,park_thesis_2012}) 

\begin{align}
\label{eq:masscons}
&\rhoin \vin = \rho_{\rm b} \vrel \, ,\\[5pt]
\label{eq:momcons} 
&\rhoin(\vin^2 +\csin^2) =  \rho_{\rm b} (\vrel^2 +\cs^2) \,.
\end{align}    

This system of equations admits two solutions. These describe a D-type (dense type) ionization front, which forms at low velocities ($\vrel < \vD \approx \cs^2/ 2\csin$) and an R-type front (rarefied type), which forms at high velocities ($\vrel > \vR \approx 2\csin$) (the definitions of $\vD$ and $\vR$ are given in \autoref{app:PR_model}).  In the intermediate velocity range, neither applies. Instead, as $\vin$ reaches the speed of sound $\csin$, a shock front forms in front of the ionized region. The relative velocity $\vin$ remains fixed to approximately the speed of sound $\csin$, while the gas density $\rhoin$ increases with the outer velocity $\vrel$ according to Eq.~\eqref{eq:momcons} (Eq.~\eqref{eq:masscons} no longer holds hold in the presence of the shock front). The explicit expression for the PR accretion rate is also given in \autoref{app:PR_model}.

In the high-velocity regime, $\vrel \gg \csin $, the BHL and PR models converge, as the ram pressure of the ambient gas dominates over the ionization pressure. In practice, however, the BHL model is multiplied by the fudge factor $\lambda$ and hence predicts accretion rates that are lower by a factor $\lambda$.
The PR accretion rate deviates significantly from BHL for velocities lower than $\vR$, when the shock front is formed.
While the BHL rate increases sharply towards lower speeds, the PR rate decreases. In this regime, the PR rate is strongly suppressed with respect to the BHL rate. 

The sound speed $\csin$ of the ionized gas, or equivalently its temperature, determines the value of $\vR$ and hence the peak of the PR accretion rate. Its value is in principle determined by the balance between heating and cooling processes within the gas. In the PR model, it is treated as a free parameter. We consider as a benchmark value $\csin = 23 $ km/s, corresponding to a temperature of $4 \times 10^4$ K~\cite{park_accretion_2013}.
{In \autoref{sec:csin}, we} provide a detailed discussion on {how the choice of $\csin$ affects the CMB bounds}.

The BHL and PR accretion rates are shown as a function of redshift in~\autoref{fig:mdot}, for the setup described in \autoref{sec:physicalScenario} and for a few different values of $\csin$ in the PR case. The plot clearly shows the effect of the radiation feedback on the accretion rate of a PBH population during the Dark Ages and naturally motivates a careful study of the impact {of different choices of $\csin$} on the CMB bound. In particular, lower values of $\csin$ lead to higher accretion rates for the range of velocities relevant to our work (see \autoref{app:PR_model}).  The change in the behavior for low $\csin$ in the PR case that appears at high redshift reflects the functional dependence  of $\dot{M}$ on the {PBH} velocity.

\section{The role of DM mini-halos}
\label{sec:mini-halos}
If PBHs constitute only a subdominant fraction of the DM, we expect the remainder of the DM to form dense compact mini-halos around them~\cite{Mack:2006gz,Berezinsky_2013,Delos:2017thv,Adamek:2019gns,Boudaud:2021irr}. These are formed in the radiation era soon after the PBHs themselves, as the DM particles decouple from the Hubble flow under the PBH's gravitational attraction. The turnaround radius $r_\mathrm{t.a.}$ for a DM particle that decouples at time $t_\mathrm{t.a.}$ is well approximated  by $r_\mathrm{t.a.}^3 \simeq 2 G M t_\mathrm{t.a.}^2 $~\cite{2019PhRvD.100b3506A}. 
In the radiation era $t \propto a^2 $, so this process leads to the formation of a density spike with a steep profile, $\rho \propto {r^{-\alpha}}$, {with $\alpha=-9/4$}. After equality, the halo keeps growing by self-similar secondary infall, expected to lead to the same radial dependence~\cite{1985ApJS...58...39B, Carr_2021}. The subsequent dynamical evolution of the DM particles in the spike can alter the profile~\cite{Eroshenko_2016, Carr_2021, Boudaud_2021}, but the simple $\alpha=9/4$ power-law remains nevertheless a good description for the range of PBH masses we consider. This profile has also been confirmed by numerical simulations~\cite{2019PhRvD.100b3506A,Serpico:2020ehh}. 
The DM mini-halos create an additional gravitational potential around the BH, which is expected to boost the accretion rate~\cite{Mack:2006gz,Serpico:2020ehh}. In particular, it was shown in Ref.~\cite{Serpico:2020ehh} that DM mini-halos can have a dramatic impact on the CMB bound on \fPBH, making it stronger by orders of magnitude. That analysis was carried out assuming BHL accretion;
in this work, we will extend this modeling to  assess the role of DM mini-halos in the context of the PR model. 

\subsection{Analytical modelling}
As discussed in \autoref{sec:accretion:models:PR}, the PR model is based on the description of a BHL accretion problem taking place within an ionized region around the BH. 
This consideration allows us to apply a similar treatment for the inclusion of DM mini-halos to that employed in~\cite{2016ApJ...818..184P,Serpico:2020ehh} for BHL. This  consists in replacing the Bondi radius $\rB$ in Eq.~\eqref{eq:bondiradius} with an effective radius $\rBeff$, which accounts for the gravitational pull of the DM halo. We write the accretion rate as 
\begin{equation}
\label{eq:MdotDM}
\Mdot  = 4 \pi \rho_{\rm b} \, \veff \, (\rBeff)^2  \, .
\end{equation}
For the BHL model, $\veff^\mathrm{BHL} =  {\left(\cs^2 + \vrel^2\right)^{1/2}}$. In the case of the PR model, the relevant effective velocity must be defined within the ionized region, hence we set $\veff^\mathrm{PR} =  {(\csin^2 + \vin^2 )^{1/2}}$. Given an effective velocity, the effective Bondi radius $\rBeff$ is the radius that satisfies the following equation 
\begin{equation} 
\label{eq:criticalradius1}
\veff ^2 = \dfrac{G M}{r} - \phi_\mathrm{h}(r)\,,
\end{equation}
{where $\phi_\mathrm{h}(r)$ is the gravitational potential generated by the DM mini-halos.}  In the PR case, this approach is self-consistent as long the effective radius obtained in this way does not exceed the size of the ionized region. 
We verify that this is true for the parameters relevant to this work, {{assuming, as estimated in~\cite{2020MNRAS.495.2966S}, that the size of the ionized region is around $10^2$ times larger than the Bondi radius computed within the ionized region. 

The {square root of the} right-hand side of Eq.~\eqref{eq:criticalradius1} 
represents an escape velocity, {which depends on $r$}. {For $r < \rBeff$}, this {velocity}  is larger than the effective velocity $\veff$ {that characterizes the gas}, and the gas can be captured and accreted by the PBH. Conversely, when $\veff$ is larger than this quantity, due to either a large sound speed or large flow velocity of the gas around the PBH, the gas is able to escape the gravitational field of the PBH. 

We model the DM mini-halos with a power law density profile $\rho_{\rm h}(r)$ and a sharp cutoff at the halo radius $r=\rh$ 
\begin{equation} 
\rho_{\rm h}(r) = \rho_{\rm h , 0} \left( \dfrac{r}{\rh} \right)^{-\alpha} \,,\quad r< r_{\rm h}.
\end{equation}
Motivated by the discussion at the beginning of this section, we set the power law index to $\alpha = 9/4$. The halo radius and the normalization constant depend on the PBH mass and evolve with redshift. Adopting the same  model {as the one} employed in~\cite{Serpico:2020ehh}, we set the total halo mass $\MH$ and halo radius $\rh$ respectively to
\begin{align}
\begin{aligned}
&\MH \simeq \dfrac{3000}{1+z} M \, ,\\
&\rh \simeq  58 \, \mathrm{pc} \, (1+z)^{-1} \left( \dfrac{\MH}{\Msun} \right)^{1/3}\, .
\end{aligned}
\end{align}
The gravitational potential generated by the DM mini-halos is then given by
\begin{equation}
\label{eq:potential}
\phi_\mathrm{h}(r) 
= \quad
\begin{cases}
    \quad
    \dfrac{G \MH}{\alpha - 2}\, \left( \dfrac{3 - \alpha}{ \rh}  - \dfrac{r^{2-\alpha}}{\rh^{3-\alpha}} \right)\, \; ,  \qquad
    & \, r < \rh \; , \\[10pt]
    \quad  
    - \dfrac{G \MH}{r}\ \; , \qquad
    & \, r \geq \rh \; \; .
\end{cases}\\
\end{equation}
Combining Eqs.~\eqref{eq:criticalradius1}~--~\eqref{eq:potential}, we obtain an equation for the effective Bondi radius for a given value of the effective velocity 
\begin{equation}
\label{eq:vsquared_effective}
\veff^2 = \quad
\begin{cases}
    \quad
    \dfrac{G M }{\rBeff} +  \dfrac{G \MH}{(\alpha-2)}\, \dfrac{1}{\rBeff} \left[ \left(\dfrac{\rBeff}{ \rh}\right)^{3 - \alpha} - \left(3 - \alpha \right) \dfrac{\rBeff}{\rh} \right]\,  \; , \qquad
    & \, \rBeff < \rh \; , \\[10pt]
    \quad  
       \dfrac{G M }{\rBeff} + \dfrac{G \MH}{\rBeff}  \; , \qquad
    & \, \rBeff \geq \rh \; \; .
\end{cases}
\end{equation}
From here we can obtain the effective Bondi radius for the BHL and PR cases, setting $\veff= \veff^\mathrm{BHL}$ and $\veff= \veff^\mathrm{PR}$, respectively. We solve Eq.~\eqref{eq:vsquared_effective} numerically for both models. Having obtained the effective Bondi radius, we compute the enhanced accretion rate through Eq.~\eqref{eq:MdotDM}.

\subsection{Impact on the accretion rate: BHL and PR scenarios}
In the BHL case, we find that the effective Bondi radius increases significantly towards low redshifts, becoming larger than the Bondi radius {in the absence of DM mini-halos} by orders of magnitude.  
In contrast, we find that when the PR model is considered, 
only a slight increase in the effective radius occurs, and only for large PBH masses. In most cases, the growth is negligible. 
\begin{figure}[t!]
\includegraphics[width=\columnwidth]{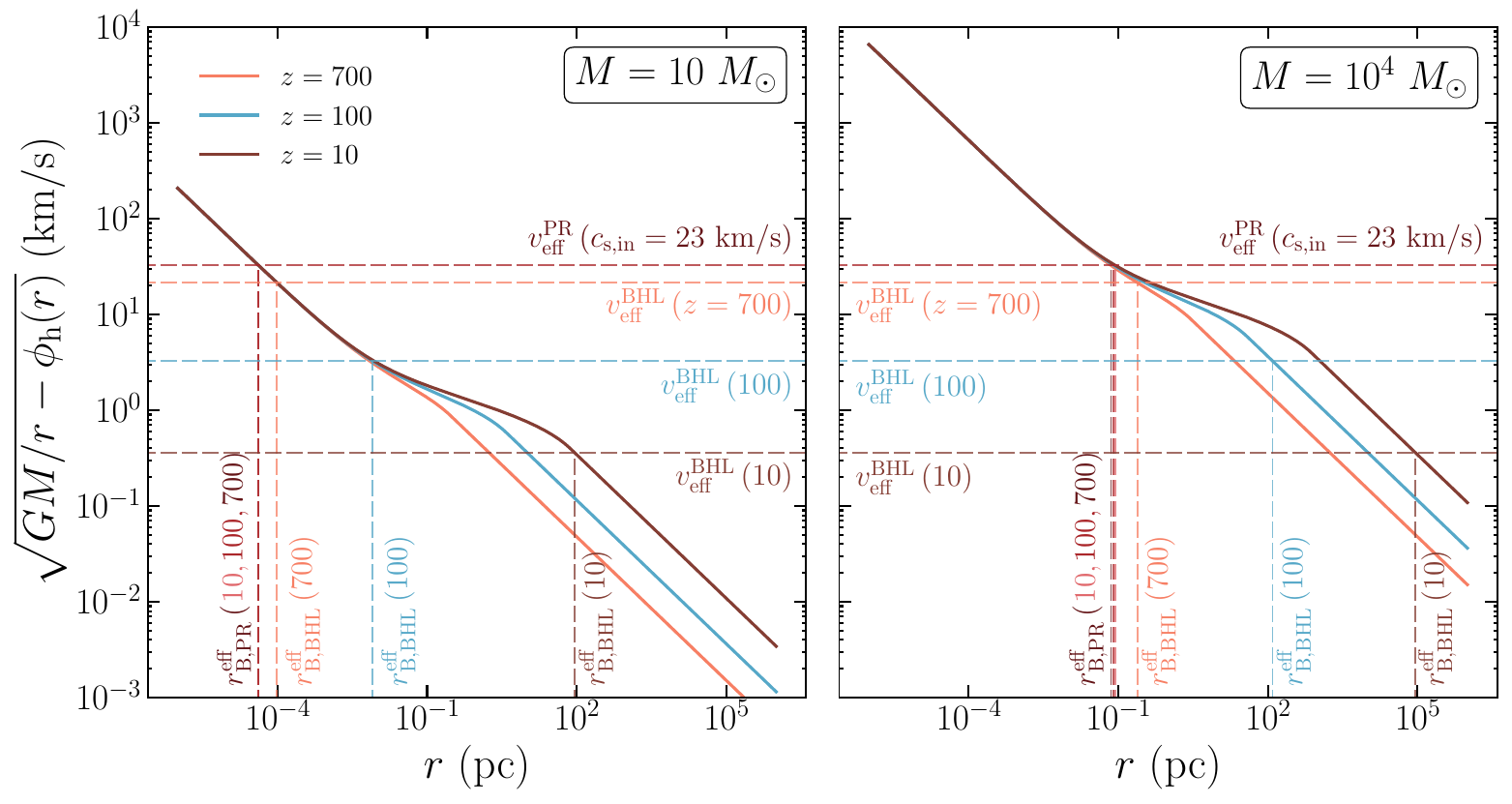} \caption{Graphical representation of the right-hand-side of Eq.~\eqref{eq:criticalradius1}. The solid lines trace the sum of the total ({PBH} plus {DM mini-}halo) potential at different redshifts.  The horizontal lines show the values of the effective velocity $\veff$ within the physical setup described in \autoref{sec:physicalScenario}, for a fixed $\Lambda $CDM cosmology with $\fPBH = 10^{-15}$. For the BHL model, $\veff $ varies with $z$. Within the PR model, it remains {almost} constant. The intersection between the horizontal lines and the curves describing the potential determines the value of the effective Bondi radius $\rBeff \left(z \right) $, denoted here by $r^\mathrm{BHL}_\mathrm{eff}$ and $r^\mathrm{PR}_\mathrm{eff}$ for the BHL and PR models, respectively.   }
\label{fig:potential}
\end{figure}

The reason for this strikingly different behavior is illustrated in \autoref{fig:potential}. 
We show the evolution with redshift of the total potential $-\phi_\mathrm{tot} = -\phi_\mathrm{BH}-\phi_\mathrm{h}$, with $\phi_\mathrm{BH} = - GM/ r$, and of the effective velocity $\veff$. The effective Bondi radius is given by the intersection between these curves, according to Eq.~\eqref{eq:criticalradius1}.

Two factors contribute to the increase of the effective Bondi radius towards low redshifts: on the one hand, the increasing volume of the DM spike; on the other hand, the decreasing value of the effective velocity.
As the DM potential builds up in the outskirts, the central region ($r \lesssim 1$ pc, depending on the PBH mass) remains dominated by the gravitational field of the PBH. At early times, high temperatures and large  baryon gas-DM {relative} velocities $\vrel$ (see \autoref{sec:physicalScenario}) both contribute to a high $\veff$. Consequently, only the central PBH-dominated region is involved in the accretion process. At later times, the temperature decreases, and the relative velocity is redshifted. In the BHL case, this implies a decreasing $\veff$, and the inclusion of a progressively larger portion of the DM-halo within the radius relevant for accretion. Especially for large masses, the contribution of the halo to the gravitational potential  eventually comes to dominate the accretion process, as the effective Bondi radius extends beyond the size of the DM mini-halo.

\autoref{fig:MdotWHalo} shows the ratio between the accretion rates with and without DM mini-halos. In the case of BHL accretion (left panel), the effect is dramatic, especially at low redshifts: the accretion rate is boosted by up to five orders of magnitude.
The transition between two regimes, a PBH-dominated regime at high redshift and a halo-dominated regime at low {redshift}, can be clearly appreciated in the figure: at low redshifts, the DM-halo is entirely enclosed by the effective Bondi radius, and the increase of the accretion rate is driven by its growth.

The picture changes in the PR scenario. In this case, the relevant effective velocity is determined by the temperature and velocity of the gas \textit{within the ionized region} surrounding the BH. The much higher temperature of the ionized gas maintains $\veff$ high. Furthermore, the relative velocity of the ionized medium with respect to the PBH remains constant (in the intermediate velocity regime, the gas velocity is set to by $\vin = \csin$). Consequently, $\veff$ does not decrease with time. The only factor contributing to the variation of the Bondi radius is the growth of the DM halo. However, as shown in  \autoref{fig:potential}, the high value of $\veff $ significantly restricts the region relevant for accretion. The central part of the potential is dominated by the PBH potential, and hence the DM spike {build-up} has a negligible effect (a small variation in $\rBeff$ due to the growth of the halo can be appreciated for $\MBH = 10^4 M_\odot$ at $z \sim 10 $).
\begin{figure}[t!]
\includegraphics[width =\columnwidth]{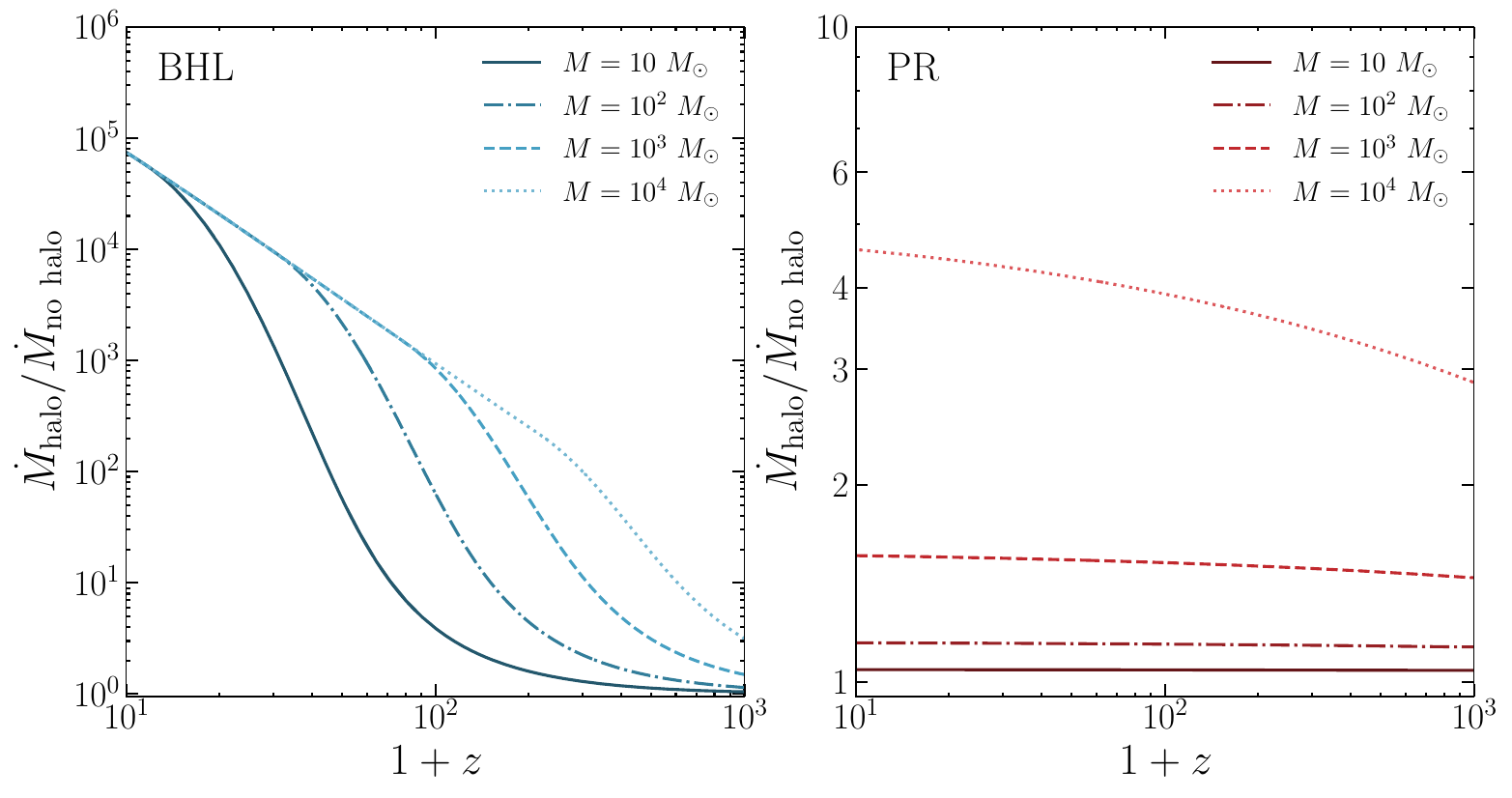}
\caption{Ratio between the accretion rates with and without DM mini-halos around the PBH, for the BHL (left) and PR (right) models. Here we assume the same fixed cosmology as described in the caption of \autoref{fig:mdot}.} 
\label{fig:MdotWHalo}
\end{figure}

The right panel of \autoref{fig:MdotWHalo} shows that, when the PR model is considered, the ratio between the accretion rates remains of order one for all redshifts: the high effective velocity prevents the DM mini-halo from contributing significantly to the accretion process. 

\subsection{General remarks}
While we have obtained this result in the context of a specific model, our analysis leads us to a more general conclusion. 
We recognize that the high temperature (sound speed) of the local gas is a sufficient condition for the DM mini-halos to play a marginal role in the accretion process. 
For typical values of the ionized gas {temperature}, $T \sim \mathcal{O} (10^4)$, accretion is mostly confined to the volume dominated by the {PBH} 
 potential, and the ratio between the accretion rate with and without  DM mini-halos is never larger than a factor of a few (for the values of the PBH mass considered here).
This result has an intuitive interpretation: the high velocity dispersion of the heated gas particles reduces the effective cross section for the accretion process. This process remains confined to a small volume around the {PBH}, where the gravitational field is dominated by the {PBH} potential.

While the PR model provides a simple prescription to account for the  local increase in temperature around the PBH, this is expected to occur in general, as a consequence of the accretion of gas and corresponding energy release. The energy injection is necessarily bound to have some degree of locality and hence to cause the local effective velocity to rise.
This discussion suggests that CMB constraints presented in the literature, which to the best of our knowledge did not take this into account, may have significantly overestimated the impact of DM mini-halos on the bound.

\section{Constraining the PBH abundance with CMB physics}
\label{sec:PhysicsBound}
In this section, we describe in detail how the accretion of gas onto PBHs is expected to modify the anisotropies of the CMB, and how we can use this observable to constrain the PBH abundance.
In particular, we discuss the phenomena governing the emission of radiation from the accreting gas and its deposition into the plasma, as well as the associated modelling uncertainties. We discuss the impact of this energy injection on the hydrogen ionization fraction and on the observed CMB temperature power spectrum. Finally, we present some details of our statistical analysis.

\subsection{Energy injection from accreting matter}
\label{sec:Einjection}
The total energy injection into the medium per unit volume is given by
\begin{equation}
\left. \frac{\mathrm{d}^2 E}{\mathrm{~d} V \mathrm{~d} t} \right|_{\mathrm{inj}}=L n_{\mathrm{PBH}} =L \fPBH \frac{\rho_{\mathrm{DM}}}{M}\ ,
\end{equation}
where $L$ is the bolometric accretion luminosity of a PBH of mass $M$, which is a function of the accretion rate $\dot{M}$.
Accretion is a very powerful energy converter. The bolometric luminosity is related to the total influx of rest-mass energy of the accreted material via the $\epsilon$ parameter that measures the {\it radiative efficiency}:
\begin{equation}
L = \epsilon {\dot{M}c^2}  \ .
\end{equation}
{Here,} $\epsilon$ is, in general, a function of $\dot{M}$, and typically  takes the functional form of a power law $\epsilon(\dot{M}) \propto \dot{M}^a$, where $a$ and the normalization can vary significantly depending on whether or not an accretion disk is formed, and on the type of accretion disk.

The basic criterion used to assess whether an accretion disk can form is to estimate the angular momentum of the baryons: if it is large enough to keep the gas in Keplerian rotation well beyond the innermost stable orbit, then a disk can form~\cite{1976ApJ...204..555S}. Following~\cite{poulin_cmb_2017}, the condition for disk formation can be written as
\begin{equation}
\fPBH \frac{M}{M_{\odot}} \ll \left( \frac{1+z}{730} \right)^3 \ .
\end{equation}
We make the assumption that a disk \textit{always} forms.\footnote{We have verified {\it a posteriori} that this is mostly the case for the range of PBH masses considered in this work for allowed values of $\fPBH$ (including those ruled out by non-CMB bounds as shown in \autoref{fig:mainPlot}) {and} in the redshift range that has the most constraining power, i.e., typically $z \sim \mathcal{O}(100-1000)$.} We relax this assumption in \autoref{sec:conservative} to derive a conservative bound.

The complex processes that shape the transport of energy and angular momentum in the disk and the associated non-thermal emission represent an {important} source of uncertainty when assessing the CMB bound. 
In accretion disks, the angular momentum of the rotating matter is gradually transported outwards by stresses (related to turbulence, viscosity, shear, and magnetic fields), and matter flows inwards. Energy is typically transferred to the electrons, which are responsible for the non-thermal cooling via synchrotron, bremsstrahlung, and inverse Compton, and is ultimately radiated away. The accretion scenario that we are considering is characterized by low values of the accretion rate, well below the Eddington value. In this situation, the accretion flow is typically not dense enough to guarantee efficient coupling between ions and electrons, as is instead the case for {\it thin disks}~\cite{1973A&A....24..337S}. Hence, the dissipated energy is not efficiently radiated away by the leptonic component and is advected into the central black hole. The accretion disk that corresponds to this scenario is hot and geometrically thick, and is usually called {\it Advention-Dominated Accretion Flow} (ADAF). The $\epsilon$ parameter in the ADAF model is linearly suppressed for smaller accretion rates.

The physics behind the suppression of the radiative efficiency is typically captured by a parameter called $\delta$, which quantifies the fraction of turbulent energy that heats the electrons directly. 
Magneto-hydro-dynamic (MHD) turbulence is the main physical process ultimately responsible for the transfer of energy and angular momentum in the accretion flow: turbulent energy is dissipated at small scales via kinetic damping mechanisms that generally transfer energy to the ions and electrons at different rates, with a preference to the heavy component. Hence, this parameter was initially believed to be very small, $\mathcal{O}(10^{-3})$ in early works on ADAF modeling. However, further investigation has demonstrated that the electrons can indeed receive a comparable fraction of turbulent heating to that of the ions due to a variety of mechanisms including magnetic reconnection~\cite{Arzamasskiy_2019,2021ApJ...916..120C,2019PhRvL.122e5101Z}.
In this work we do not aim at modeling the complex physics that shapes these processes. Instead, we follow the approach of~\cite{Xie_2012}: we bracket the associated uncertainties by treating $\delta$ as a free parameter. For different reference values of $\delta$, the authors solve a set of equations that describe a two-fluid ADAF and compute both its dynamical structure (temperature and density profile) and the non-thermal emission.

The main output of these calculations is to define the behavior of the radiative efficiency $\epsilon$ as a function of the accretion rate $\dot{M}$, which has the functional form 
\begin{equation}\label{eq:epsilon}
\epsilon (\dot{M}) = \epsilon_0 \left(\frac{\dot{M}}{0.01 \dot{M}_{\rm edd}}\right)^{a},
\end{equation}
where $\dot{M}_{\rm edd}$ is the Eddington accretion rate. The parameters $\epsilon_0$ and $a$ characterizing the model take a piecewise functional form explicitly dependent on the accretion rate,  and are defined in table I of~\cite{Xie_2012}. For typical values of the accretion rate in the PR model, using the benchmark of $\delta = 0.1$, we have $\epsilon_0 = 0.12$ and $a = 0.59$; this possibility, as well as the other values that we consider in this paper are detailed in \autoref{tab:deltas}.  
\begin{table}[t]
\centering
\begin{tabular}{|c|c|c|c|}
\hline  $\delta$ & $\dot{M}_{\text {net }} / \dot{M}_{\text {Edd }}$ Range & $\epsilon_0$ & $a$ \\
\hline \hline  & $<2.9 \times 10^{-5}$ & 1.58 & 0.65 \\
$0.5$ & $2.9 \times 10^{-5}-3.3 \times 10^{-3}$ & 0.055 & 0.076 \\
& $3.3 \times 10^{-3}-5.3 \times 10^{-3}$ & 0.17 & 1.12 \\
\hline  & $<9.4 \times 10^{-5}$ & 0.12 & 0.59 \\
$0.1$ [benchmark] & $9.4 \times 10^{-5}-5.0 \times 10^{-3}$ & 0.026 & 0.27 \\
& $5.0 \times 10^{-3}-6.6 \times 10^{-3}$ & 0.50 & 4.53 \\
\hline  & $<7.6 \times 10^{-5}$ & 0.065 & 0.71 \\
$10^{-3}$ & $7.6 \times 10^{-5}-4.5 \times 10^{-3}$ & 0.020 & 0.47 \\
& $4.5 \times 10^{-3}-7.1 \times 10^{-3}$ & 0.26 & 3.67 \\
\hline
\end{tabular}
\caption{The piecewise power-law fitting formulae giving the radiative efficiency parameters needed to compute Eq. \eqref{eq:epsilon}, taken from \cite{Xie_2012}. }
\label{tab:deltas}
\end{table}
{The function in Eq.~(\ref{eq:epsilon})} captures all the information that is needed for the problem of energy injection into the {inter-galactic medium (IGM)} that is of interest for our current work.  
We remark that our benchmark choice implies a larger radiative efficiency compared to the (very conservative) scenario of spherical accretion (as investigated by~\cite{AliHaimoudKamionkowski2017}). On the other hand, it features a lower efficiency compared to the thin-disk case, which {predicts} a constant value of $\epsilon \simeq 0.1$~\cite{1973A&A....24..337S}. 

\subsection{Energy deposition}
\label{sec:Edeposition}
It has been shown that energy injections during the dark ages are not necessarily deposited into the medium \textit{on the spot},\footnote{On-the-spot refers to energy being absorbed by the medium at the same redshift as it was emitted.} and rather can be deposited at later times. The energy deposition functions, $f_c \left( z , X_e\right) $, quantify the amount of injected energy that is deposited at redshift $z$ for a channel $c$, where the most relevant channels are the heating, ionization, and excitation of atoms. Equivalently, $f_c \left( z, X_e \right) $ can be decomposed into an  injection efficiency function  $f_{\mathrm{eff}} \left( z \right)$, and a deposition fraction $\chi_c \left(  z , X_e \right)$.  These recipes for computing the redshift dependence of energy injection and deposition are quantified through the relation
\begin{equation}
 \left. \frac{\mathrm{d}^2 E}{\mathrm{~d} V \mathrm{~d} t} \right|_{\mathrm{dep,c }} = f_c \left( z, X_e \right)  \left. \frac{\mathrm{d}^2 E}{\mathrm{~d} V \mathrm{~d} t} \right|_{\mathrm{inj}} = f_{\text {eff }} \left( z\right) \chi_c \left(  z , X_e \right) \left.\frac{\mathrm{d}^2 E}{\mathrm{~d} V \mathrm{~d} t}\right|_{\text {inj }},
 \label{eqn:dep_fn_defn}
\end{equation}
where the first equality makes use of the energy deposition {functions, $f_c$,} and the second uses the deposition fractions, $\chi_c$. Both methodologies are equivalent at the level of Eq.~(\ref{eqn:dep_fn_defn}). However, they correspond to two different approaches for computing {the energy deposition} that have been adopted in the recent literature~\cite{Galli2013, slatyer_indirect_2016}.
We quantify the effect of these {two} approaches on the bound in \autoref{sec:resultsAstro}.
\begin{itemize}
\item The {\it energy deposition fraction} prescription has been recently used to compute CMB bounds for PBHs using PR accretion~\cite{Facchinetti2022}, where the authors used the $f_{\mathrm{eff}} (z)$ computed analytically in~\cite{AliHaimoudKamionkowski2017} combined with the deposition fractions tabulated from~\cite{Galli2013}. We refer to this as the analytic case.
\item Alternatively, authors have also recently used the {\it energy deposition function} prescription, where the energy deposition functions $f_c (z)$ are calculated for a given energy-differential luminosity spectrum $L_\omega$, using the following expression:
\begin{equation}
f_c (z, X_{e}) = H(z) \frac{\int \frac{\mathrm{d} \ln \left(1+z^{\prime}\right)}{H\left(z^{\prime}\right)} \int T\left(z^{\prime}, z, \omega\right) L_\omega \mathrm{d} \omega}{\int L_\omega \mathrm{d} \omega} \ ,
\label{eq:f_c integral}
\end{equation}
where $L_\omega$ is the energy-differential luminosity spectrum, and the transfer functions $T\left(z^{\prime}, z, \omega\right) $ are tabulated from~\cite{Slatyer:2012yq,slatyer_indirect_2016}. These transfer functions are computed numerically down to energies of $5$  $\mathrm{ keV}$, however,  some authors have suggested that they can be extrapolated down to $100\,\, \mathrm{ eV}$, and thus set the lower bound in the integrals in Eq.~\eqref{eq:f_c integral} to this value~\cite{poulin_cmb_2017,Serpico:2020ehh}. We refer to this as the extrapolated case. {Rather than extrapolating it down to 100~eV,} we cut the integration at the lower bound of their calculated value ($5$~$\mathrm{ keV}$), and we refer to this as the benchmark case. This prescription is the default setting in the numerical \texttt{DarkAges} script~\cite{stocker_exotic_2018}, which we use to perform the integral in Eq.~\eqref{eq:f_c integral}. Note that one of the key assumptions in the calculation of the transfer functions using this prescription is that the altered free electron fraction does not have a feedback effect on the energy cascade evolution~\cite{stocker_exotic_2018}. Recently, a new tool, \texttt{DarkHistory}~\cite{Liu2019}, has been developed to self-consistently take into account this feedback effect and compute $X_e$-dependent transfer functions. We have chosen not to include this in our setup, as we expect the impact on the bound to be negligible.\footnote{We compared the computed free electron fraction at the bound (i.e. $\fPBH ~ 10^{-3}$ at  $M = 10^3 \, M_\odot$ in \autoref{fig:xe}) to Figure 3 of~\cite{Capozzi2023}, and verified that the impact on the computed ionization fractions between the \texttt{DarkAges} and \texttt{DarkHistory} treatments is negligible. }
\end{itemize}

\subsection{Comparison with cosmological data}
\label{sec:CLASS}
\begin{figure}[t!]
\centering
\includegraphics[width=1.0\linewidth]{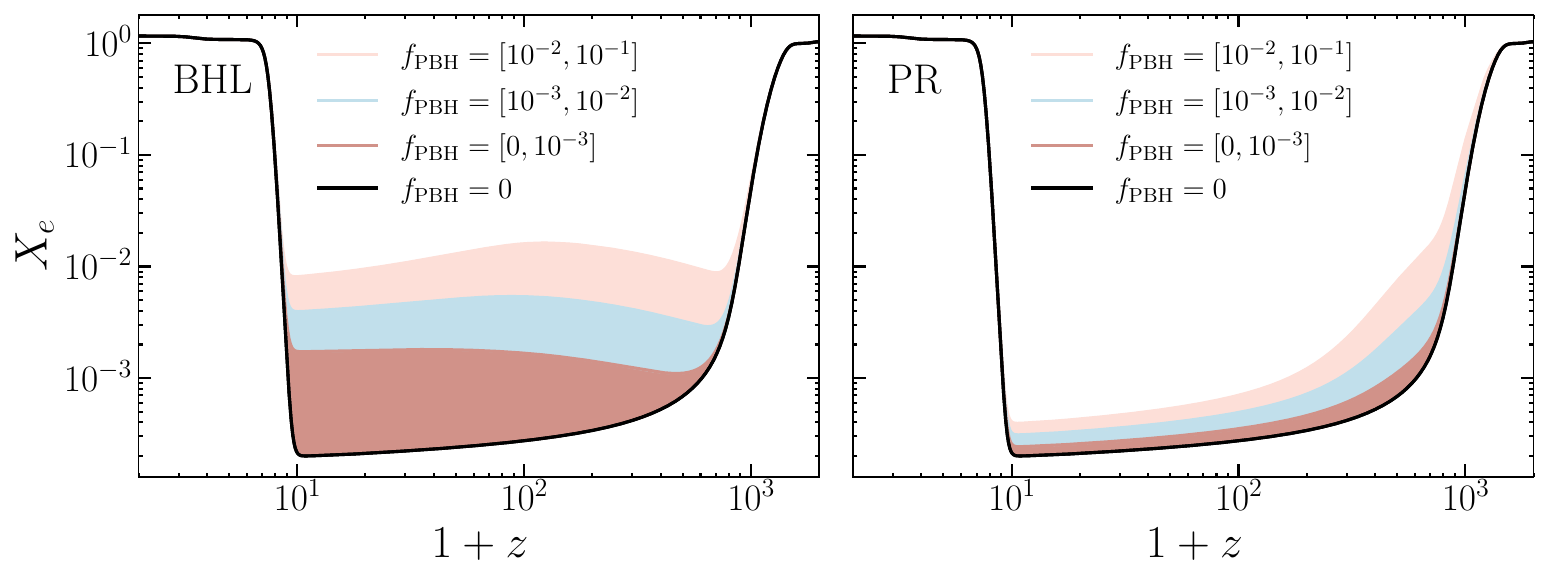}
\caption{ The free electron fraction, $X_e$, as a function of redshift, showing the effect of a monochromatic PBH population with mass $M_\mathrm{PBH} = 10^3 M_\odot$, for different PBH abundances, as labelled. The left figure is for the BHL accretion scenario, and the right figure shows the PR scenario, with $\csin = 23$~km/s. The standard scenario $\Lambda$CDM cosmology (with parameters described in the caption of \autoref{fig:mdot}) is described by the solid black line.  }       
\label{fig:xe}
\end{figure}

Once the energy deposition formalism is set, the last step consists of computing the impact of the extra energy that is injected into the medium on the ionization fraction $X_e$ and the IGM temperature $T$, and the subsequent alteration of the CMB power spectrum. 
The physical processes responsible for such alterations are discussed in detail for instance in~\cite{Peebles:2000pn,AliHaimoudKamionkowski2017}. The key point is the broadening of the visibility function towards lower redshift, which mainly implies {a} suppression of the secondary peaks due to an increase in the time available for the dissipation of acoustic oscillations~\cite{Hu:1995kot}.

\begin{figure}[t!]
\centering
\includegraphics[width=0.49\linewidth]{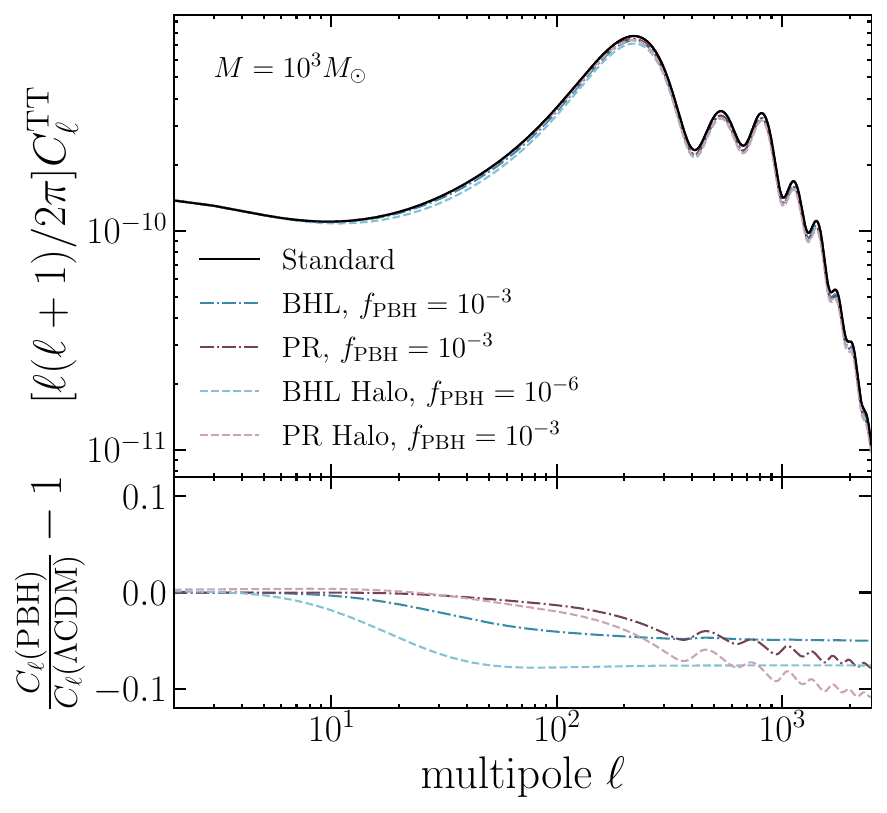}
\includegraphics[width=0.49\linewidth]{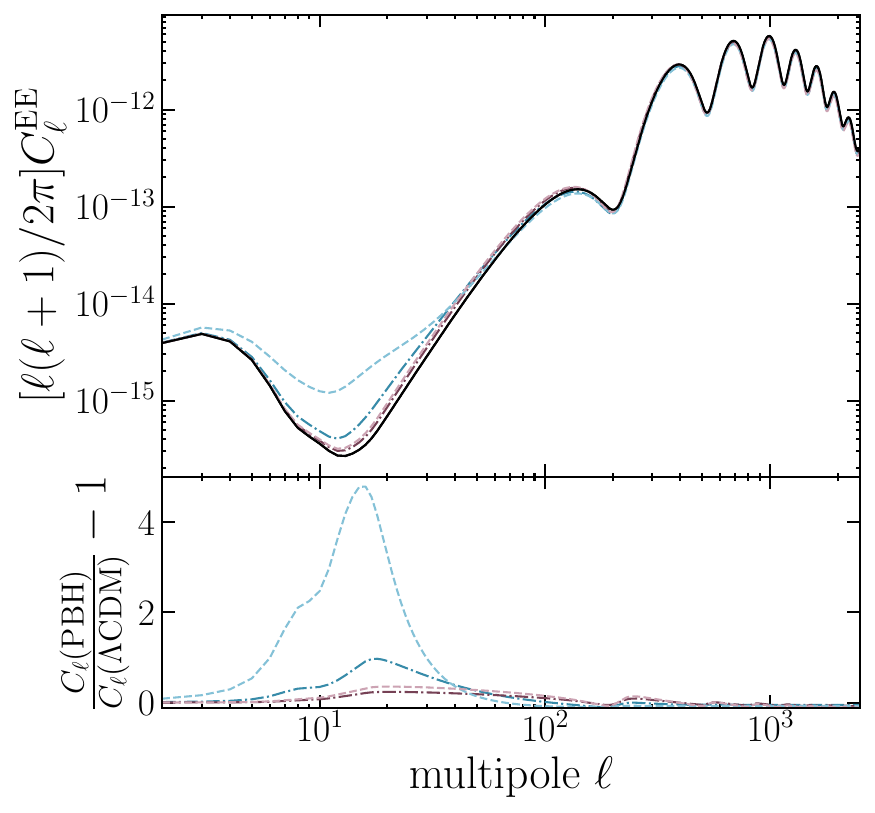}
\caption{ The impact of the accretion recipe on the CMB $TT$ (left) and $EE$ (right) power spectrum for a monochromatic population of PBHs with masses $M = 10^3  \ M_\odot$, assuming the fixed $\Lambda$CDM cosmology described in the caption of \autoref{fig:mdot}.  }
\label{fig:cls}
\end{figure}

To compute the CMB and large scale structure observables corresponding to the simulated evolution of linear perturbations from the early universe, we use the latest version of the publicly available Boltzmann solver \texttt{CLASS}~\cite{blas_cosmic_2011}. 
In particular, we implement modifications to \texttt{CLASS}, altering the thermal history of the universe to include the effect of accreting PBHs. 
We base our approach on the framework presented in \texttt{ExoCLASS}~\cite{stocker_exotic_2018}, making modifications to include our recipes. For illustrative purposes, we show in \autoref{fig:xe} and  \autoref{fig:cls} the impact of the addition of a population of radiating PBHs for different values of $\fPBH$.

This modified version of \texttt{CLASS} is then interfaced with the Bayesian analysis code \texttt{Cobaya}~\cite{Torrado2020,2019ascl.soft10019T}, to perform a Markov Chain Monte Carlo (MCMC) simulation to probe the 7-dimensional $\Lambda \mathrm{CDM} + \fPBH$ parameter space,
\begin{equation}
\Lambda \mathrm{CDM} + \fPBH \equiv \left\{ \Omega_\mathrm{b} ,\Omega_\mathrm{c},
H_0, A_\mathrm{s}, n_\mathrm{s},z_\mathrm{reio}   \right\} + \fPBH
\label{eqn:LCDM+fpbh} \ ,
\end{equation}
where $9$ masses evenly log-spaced in the range of $M_\mathrm{PBH} \in \left[ 10^1, 10^4 \right]$ $\mathrm{M}_{\odot}$ are probed. These parameters are constrained using the latest state-of-the-art likelihoods from \texttt{Planck}, \texttt{SPT}, \texttt{ACT} and \texttt{BAO}. We use flat priors for each of the cosmological parameters, described in detail in \autoref{app:mcmc_details}. In particular, we constrain the cosmological parameters \eqref{eqn:LCDM+fpbh} using \textit{Planck} 2018 high-$l$ TTTEEE, low-$l$ TT, low-$l$ EE~\cite{planck_collaboration_planck_2020}; the isotropic BAO measurements from 6dFGS at $z_{\mathrm{eff}}=0.106$~\cite{beutler_6df_2011} and from the Sloan Digital Sky Survey Data Release 7 at $z_{\mathrm{eff}}=0.15$~\cite{ross_clustering_2015}; the anisotropic BAO and growth function $f \sigma_8 (z)$ measurements made using the CMASS and LOWZ galaxy samples of BOSS DR12~\cite{alam_clustering_2017} and DR16~\cite{eboss_collaboration_completed_2021}; 2022 SPT 3g TTTEEE~\cite{balkenhol_measurement_2023}; ACT DR4~\cite{aiola_atacama_2020,choi_atacama_2020} and ACT DR6 + Planck2018 lensing~\cite{madhavacheril_atacama_2023,qu_atacama_2023,carron_cmb_2022}. We consider runs to be converged with the Gelman-Rubin criterion, when $\left| R- 1 \right| < 0.05$~\cite{gelman_inference_1992}. The bound is computed as the 95\% probability using a highest-density interval calculation~\cite{Kumar2019}.

\section{Results: The CMB bound with radiative feedback and DM mini-halos}
\label{sec:resultsBounds}
\begin{figure}[t!]
\centering
\includegraphics[width=0.98\linewidth]{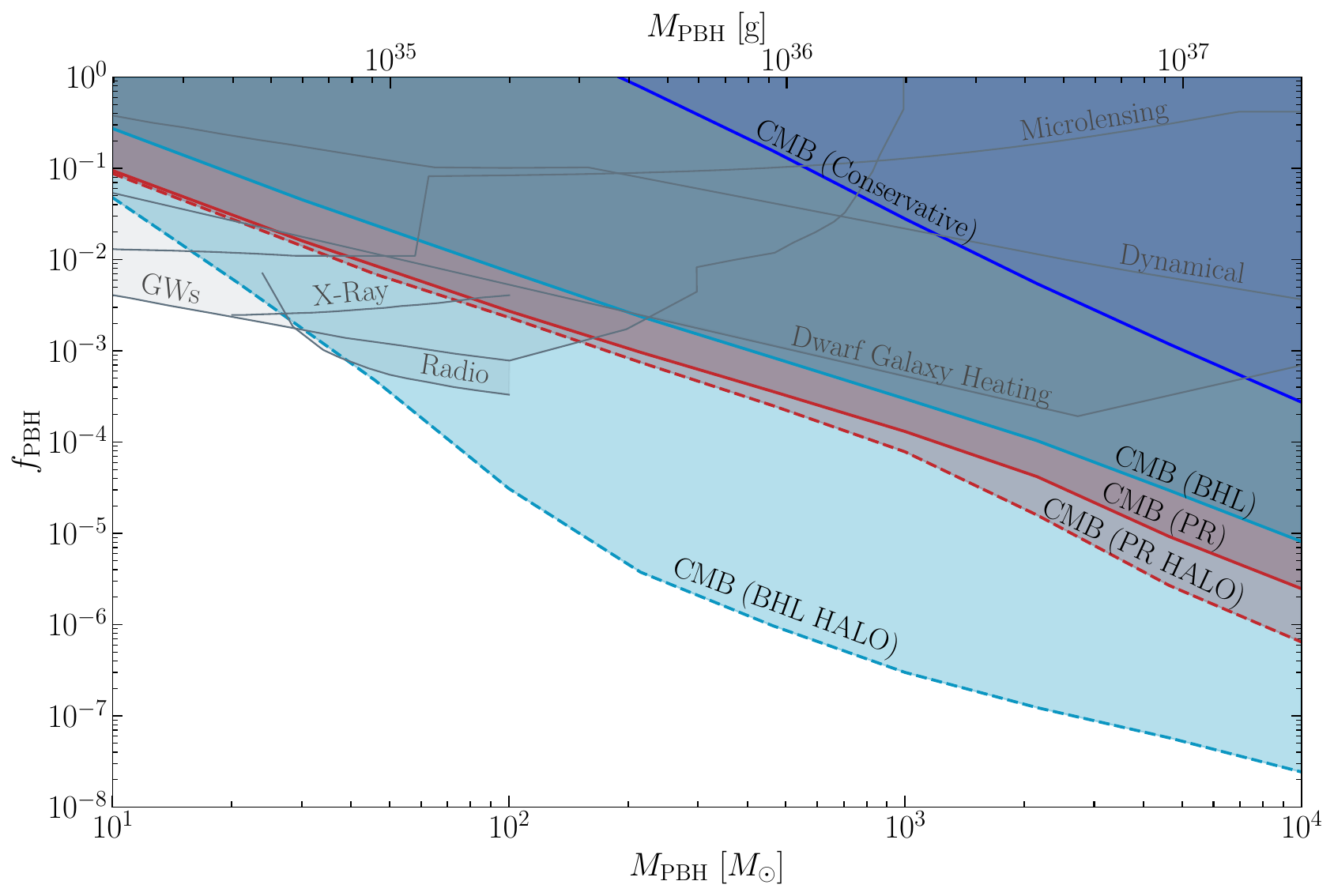}
\caption{Constraints at the 95\% probability from the posterior of $\fPBH$ -- which describes the cosmological abundance of PBHs {as a fraction of the DM abundance}, assuming {a} monochromatic PBH mass function -- obtained within different accretion recipes and for a selected range of PBH masses. In light blue is the bound assuming BHL accretion with a disk geometry {and with either} DM halos included (dashed, {``CMB (BHL HALO)'')} {or not included} (solid, {``CMB (BHL)''}). In red is the bound assuming PR accretion with a disk geometry {and with either} DM halos included (dashed, {``CMB (PR HALO)''}) or not included (solid, {``CMB (PR)''}). The dark blue line {(``CMB (Conservative)'')} {shows} the CMB accretion bound following the spherical accretion recipe of \autoref{sec:conservative}. The most stringent and relevant PBH bounds in this mass regime are also shown {from} gravitational waves~\cite{Kavanagh2018,Chen2019}, Radio and X-Ray~\cite{Manshanden2018}, Microlensing~\cite{Oguri2018,Blaineau2022,EstebanGutierrez2023}, Dynamical~\cite{MonroyRodriguez2014,Brandt2016}, and Dwarf Galaxy Heating~\cite{Lu2020}. We make use of~\cite{Kavanagh2019} for plotting. }
\label{fig:mainPlot}
\end{figure}

We now consider the benchmark values of the main free parameters under consideration (given in \autoref{tab:benchmark}), and assess the impact on the bound of the accretion model and of the inclusion of DM mini-halos.
The updated CMB bound on the abundance of PBHs, according to the assumption and the data described in the previous sections, are displayed in \autoref{fig:mainPlot} for a few selected reference scenarios: {\it (i)}~Bondi-Hoyle-Lyttleton accretion in absence of a relevant contribution of dark-matter mini halos, with the fudge factor set to $\lambda = 0.01$; {\it (ii)}~Park-Ricotti accretion in absence of DM mini-halos; {\it (iii)}~BHL accretion with DM halos; {\it (iv)}~PR accretion with DM halos.

\begin{table}[]
\centering
\begin{tabular}{|c|c|c|c|c|}
\hline Model & $\csin$ & $\lambda$ &$\delta$ & Energy deposition treatment \\
\hline \hline  
BHL & n/a & 0.01  & $0.1 $ & $f_c(z)$ computed using Eq. \eqref{eq:f_c integral} with details in \autoref{sec:Edeposition} \\ 
\hline 
PR & $23 $ km/s & n/a & $0.1$ & $f_c(z)$ computed using Eq. \eqref{eq:f_c integral} with details in \autoref{sec:Edeposition} \\
\hline 
\end{tabular}
\caption{The benchmark values of the main parameters under consideration, with full descriptions of each parameter provided in \autoref{sec:PhysicsBound}. In \autoref{sec:resultsAstro}, we show the impact on the bound when varying $\csin$ for the PR model, the parameter $\delta$, and the energy deposition treatment.}
\label{tab:benchmark}
\end{table}

\subsection{The impact of the accretion model}
Replacing the BHL model with the PR model does not significantly affect the CMB constraint, as long as the effect of DM mini-halos is ignored.
This is a highly non-trivial result, given the significant difference of the functional dependence of the accretion rate on the relative BH speed. 
The reason for this consistency is the accidental cancellation of two competing effects. On the one hand, at recombination and shortly after, the PR model provides a {\it larger} accretion rate, see \autoref{fig:mdot}. At these redshifts, the relative baryon-DM velocity tends towards values that correspond to the high-velocity regime in the PR accretion rate. In this regime, as discussed in~\autoref{sec:accretion:models:PR}, the PR model predicts accretion rates higher by a factor $\sim    \lambda$ than the BHL model. On the other hand, at lower redshifts, the velocity decreases and the PR model enters its intermediate regime, where a shock front is formed and the accretion rate becomes very strongly suppressed. The two effects partially compensate, resulting in a difference of a factor $[2.5 - 3]$ between the two bounds. In conclusion, while one may have naively expected a {\it weakening} of the bound in the PR case as a consequence of the low-velocity suppression of the accretion rate, the aforementioned cancellation actually slightly moves the bound in the opposite direction, making it more stringent. 

Varying the value of the sound speed $\csin$ can change the ratio between the PR and BHL accretion rates, as is shown in \autoref{fig:mdot}. We analyze the impact of this parameter on the PR bound in section~\autoref{sec:csin}.

\subsection{The impact of DM mini-halos}
\label{sec:resultsSpikes} 
We obtain the CMB bound including DM mini-halos following the prescription described in \autoref{sec:mini-halos}. As discussed there, this consists in computing an enhanced accretion rate expressed in terms of an effective Bondi radius. This radius defines an effective cross section for the accretion process and is obtained comparing the total potential of the BH plus the DM halo to the effective velocity $\veff$, which depends on the relative BH-gas velocity and on the gas sound speed. This is computed through  Eq.~\eqref{eq:criticalradius1} and represented in \autoref{fig:potential}.

In the context of the PR model, we assumed and verified a posteriori that the accretion process, including {in the presence of DM mini-halos}, occurs within the local ionized region around the BH (we assume this region to be around $10^2$ times larger than the Bondi radius~\cite{2020MNRAS.495.2966S}). Within this region, temperatures are much higher than those of the neutral background medium. Following the PR prescription, we set the sound speed of the ionized gas to a constant value, choosing as a reference $\csin = 23$~km/s, which corresponds to a temperature of a few tens of thousands of Kelvin. 
As already discussed in \autoref{sec:mini-halos}, the high value of the local effective velocity that follows prevents the DM mini-halos from significantly contributing to the accretion rate. This is reflected in the behavior of the constraint, which is barely affected by the inclusion of the DM mini-halos. We find an effect of at most a factor of a few at the largest PBH masses considered.

To obtain the bound in the BHL case, we perform an analysis similar to that of~\cite{Serpico:2020ehh}.\footnote{{Ref.~\cite{Serpico:2020ehh}} used results of their simulations to describe the DM profile; we rely on the analytical model. They comment in their paper that the difference in the bound is a factor $\sim 2$.}
The evolution with redshift of the effective velocity is determined by the linear velocity, which evolves as in Eq.~\eqref{eq:vrel} and by the background value of the sound speed $\cs$, which is computed by \texttt{CLASS} taking into account the energy injection from PBHs. 

Comparing the dashed lines in~\autoref{fig:mainPlot}, we observe that, when the effect of DM mini-halos around PBHs is taken into account, the constraint obtained with the PR model is weaker by up to 3 orders of magnitude than the constraint obtained with the BHL model.
As discussed in \autoref{sec:mini-halos}, the analysis carried out in the BHL case is likely to be overestimating the impact of DM mini-halos. The local increase in the gas sound speed due to the accretion process should be taken into account to correctly quantify the contribution of mini-halos to the accretion rate.

\begin{figure}[!ht]
\centering
\includegraphics[width=0.88\linewidth]{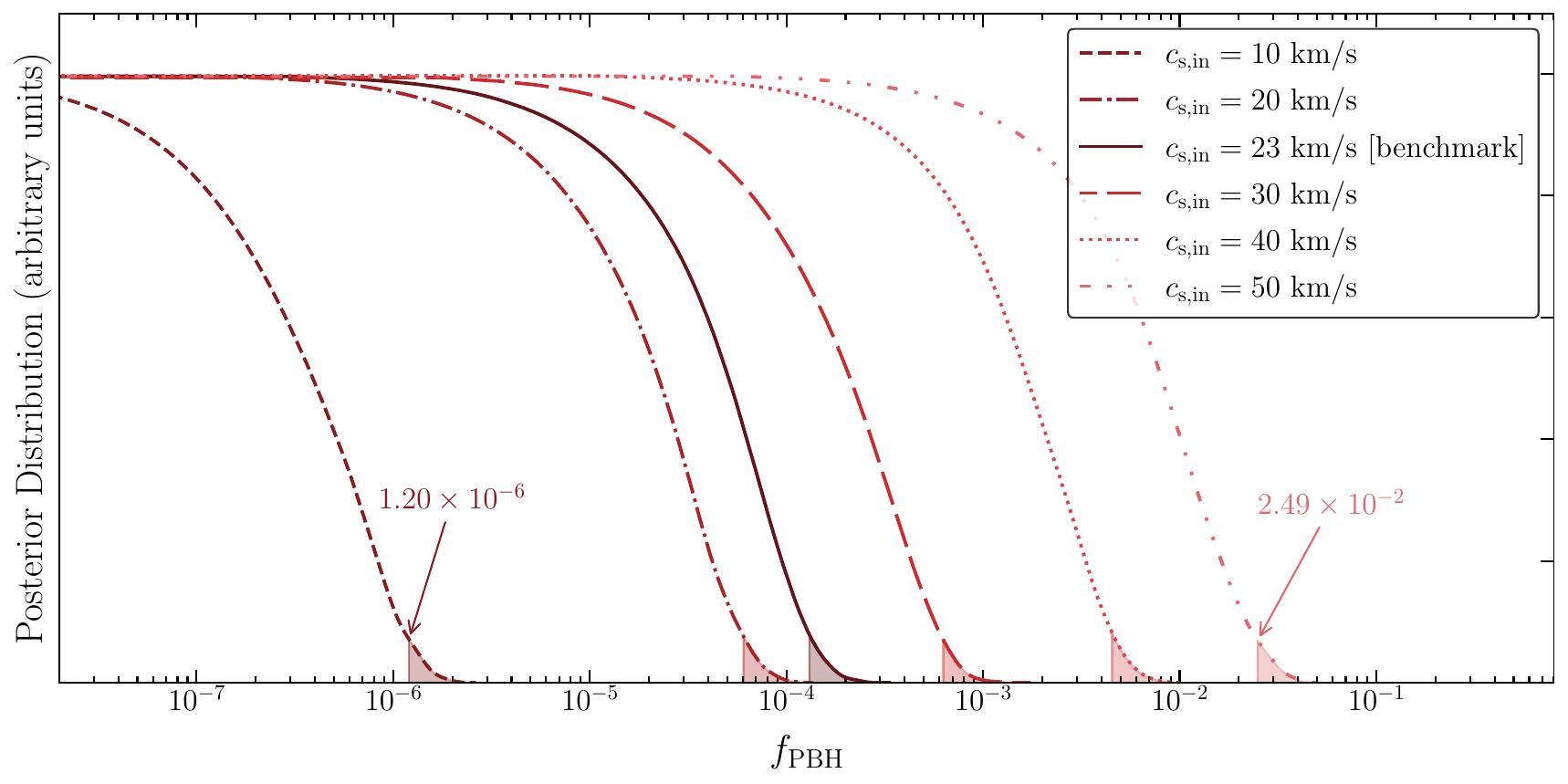}
\includegraphics[width=0.48\linewidth]{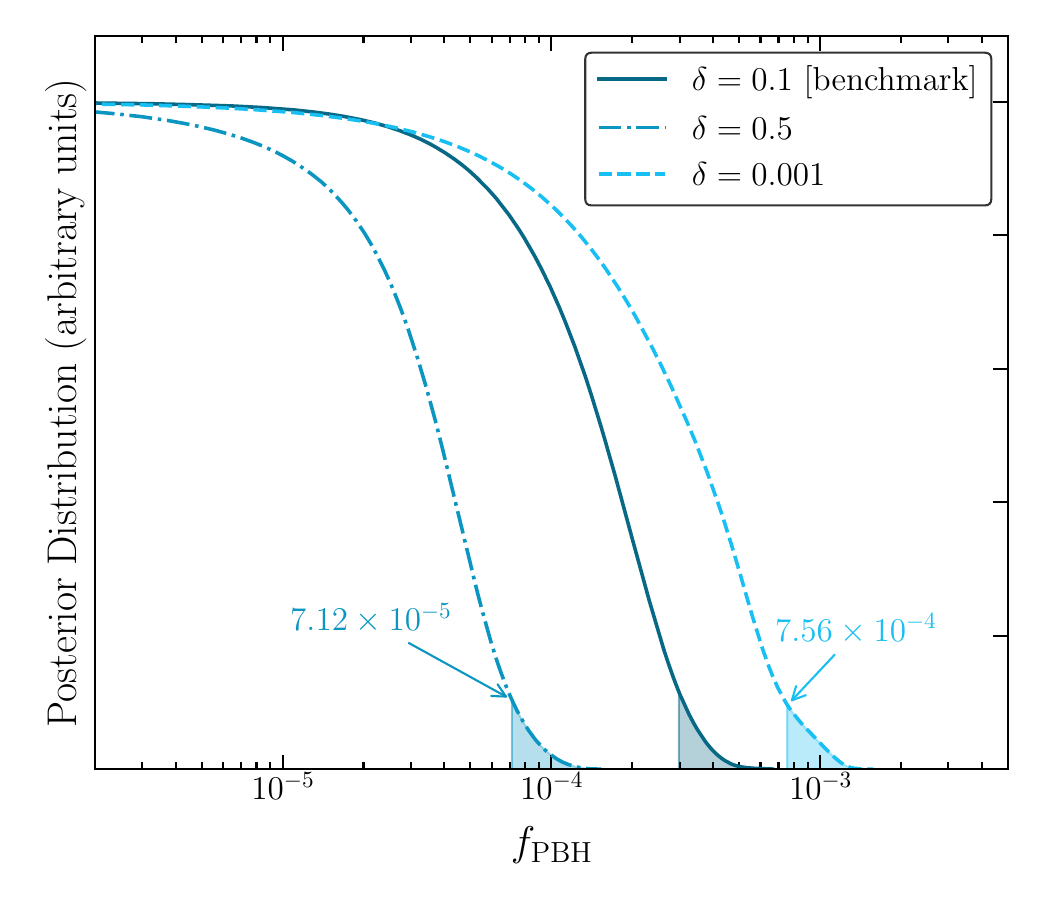}
\includegraphics[width=0.48\linewidth]{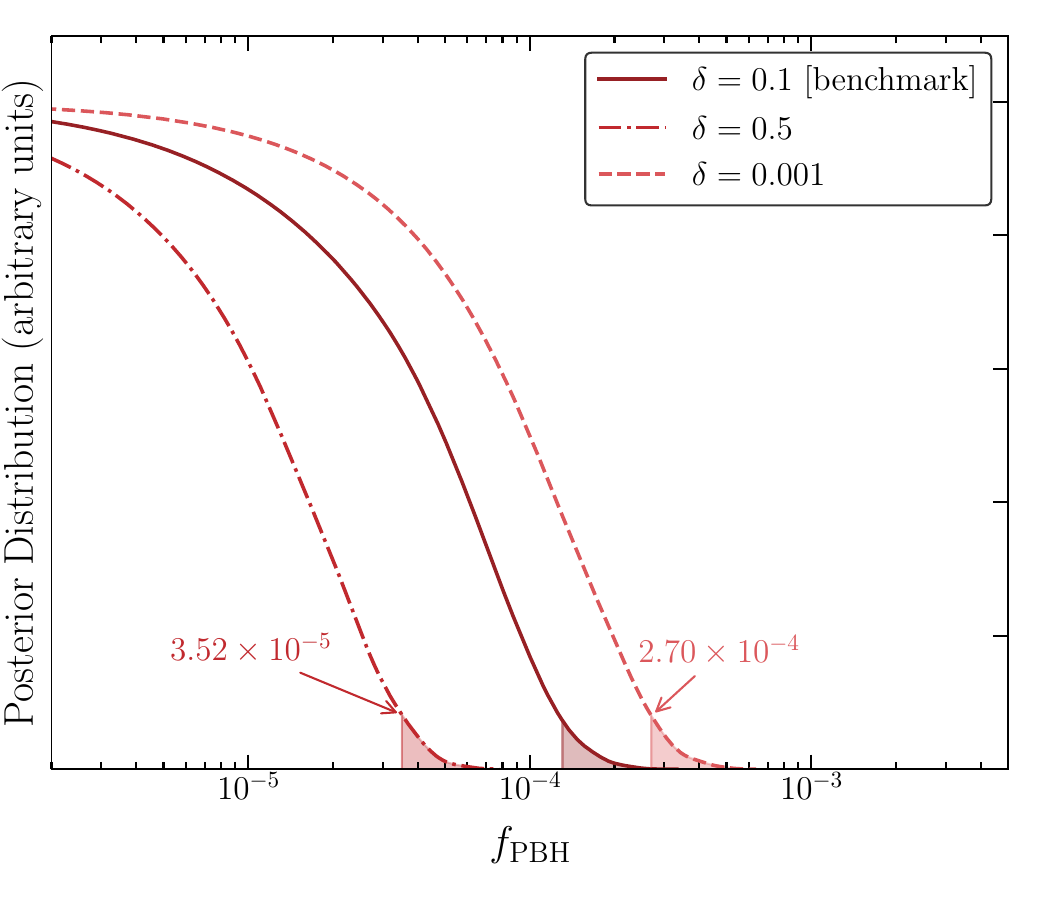}
\includegraphics[width=0.48\linewidth]{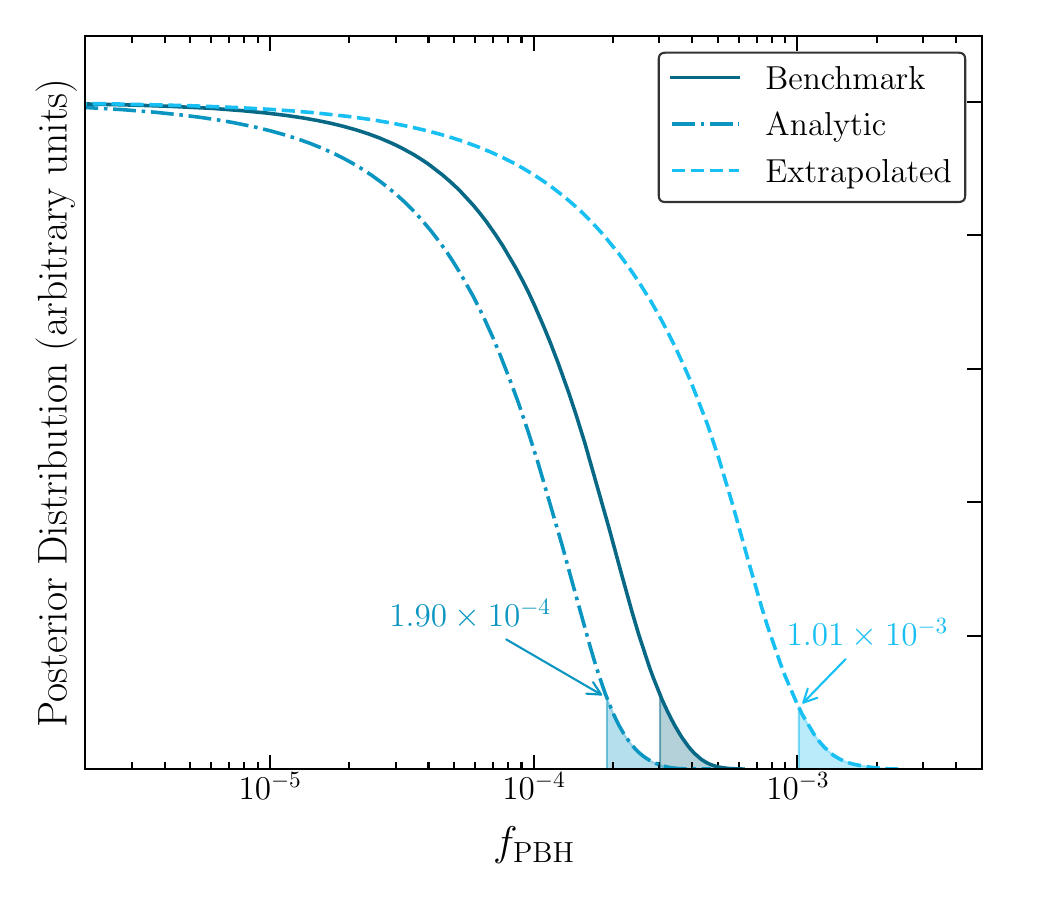}
\includegraphics[width=0.48\linewidth]{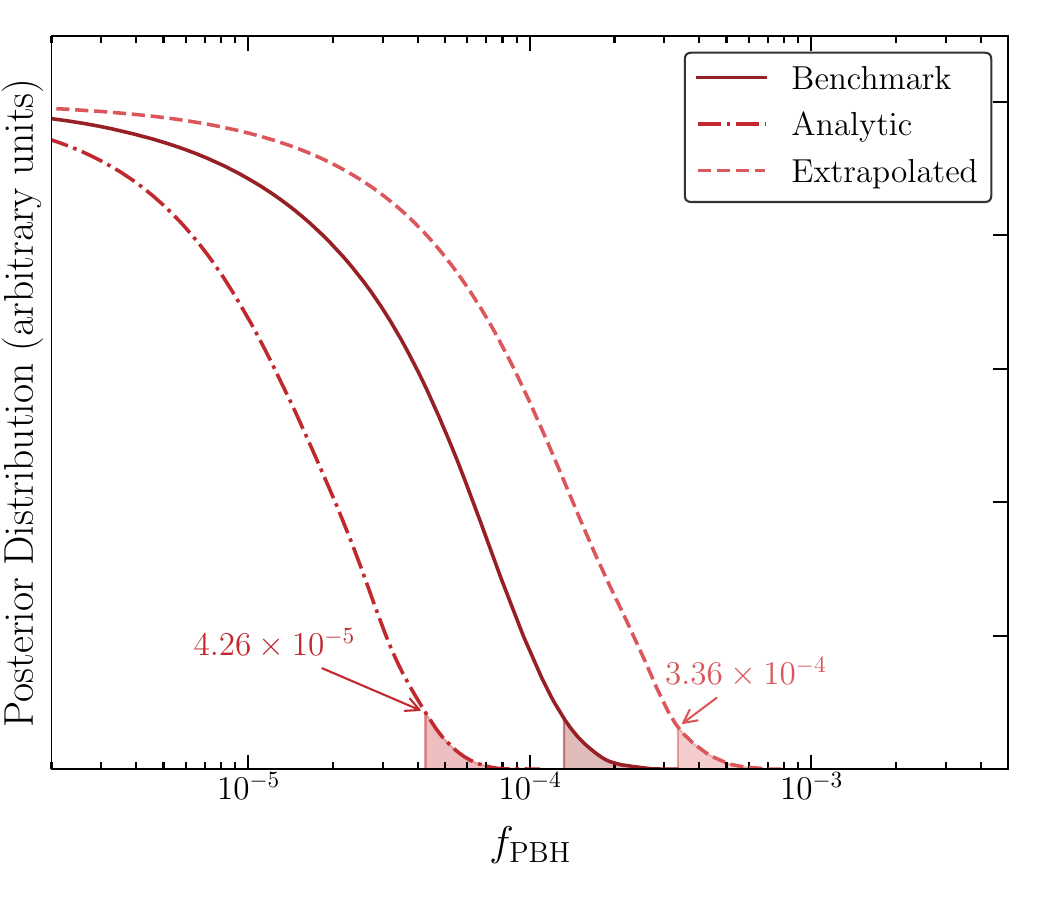}

\caption{Marginalized posterior distributions of $\fPBH$ (in arbitrary units) showing the effect of each ingredient in the accretion recipe entering in the CMB bound for a benchmark PBH mass of $ 10^3 \ M_{\odot}$. The 95\% probability region is shaded for each posterior, and the bound is labelled for the two extrema for each panel. In particular, the top panel shows the effect of  the ionized sound speed in the PR model. The middle panel shows the effect of varying $\delta$ in an ADAF accretion scenario for the BHL  (left) and the PR  (right) models (see \autoref{tab:deltas}, and Eq. \eqref{eq:epsilon} for details on $\delta$). The bottom panel shows the effect of different energy deposition function treatments for BHL (left) and PR (right).}
\label{fig:panel_figure}
\end{figure}

\section{Results: Assessment of the astrophysical uncertainties}
\label{sec:resultsAstro}
We now aim to assess the most relevant astrophysical uncertainties that affect the CMB constraints, namely:  {\it i)} the value the ionized sound speed  {\it ii)} the evaluation of the radiative output associated with the accretion disk; {\it iii)} the evaluation of the energy deposited in the IGM. 
Using a PBH mass of $10^3 \ \Msun$ as the reference value, we compute the posterior distribution function associated with the {fractional abundance of PBHs}, $f_\mathrm{PBH}$, for different values of the parameters which control the aforementioned uncertainties.
We show these posteriors in \autoref{fig:panel_figure}, shading the regions excluded at $95\%$ probability. 

\subsection{Ionized sound speed}
\label{sec:csin}
The value of $\csin$, which characterizes the temperature of the ionized region, alters the functional form of $\dot{M}$ in the PR model, varying the position of the peak as a function of the PBH speed (see~\autoref{fig:pr_vbh_cs}).
The temperature of the ionized medium is in principle fully determined by the balance between heating and cooling of the material within the ionized region. However, in the PR model it is treated as a constant free parameter. This should be regarded as an approximation for two reasons. Firstly, the temperature of the ionized medium is expected to present a gradient with distance from the BH. Secondly, the amount of heating is tied to the energy injected by the BH and is hence expected to a) be a function of the parameters that enter the accretion rate itself (PBH mass, gas density), b) depend on the efficiency of the energy emission processes. We consider that this approximation is validated by the ability of the PR model to reproduce the results of the simulations, in which the dependencies and feedback mentioned above are taken into account.

Using the results presented in~\cite{Sugimura:2020rdw} as a guide, we consider as our benchmark $\csin = 23$~km/s.
In~\cite{Sugimura:2020rdw}, simulations were performed for different values of the density and temperature of the ambient gas, as well as the BH mass. The results show that the temperature profile of the ionized gas presents order one variations with these parameters, being determined in particular by the product of gas density and BH mass.\footnote{See Figure 5 of Ref.~\cite{Sugimura:2020rdw}} Moreover, the BH luminosity was assumed to scale with the accretion rate with a slightly different recipe from the one we use here, leading to a luminosity that is lower by around a factor of ten. 

Given the uncertainty associated with this kind of modeling, it is interesting to explore how our result are affected by the value of $\csin$. 
We show our posteriors on $\fPBH$ for different choices of this parameter in the top row of \autoref{fig:panel_figure}.
{We see that the} impact on the posterior distribution function associated with the {fractional abundance of PBHs}, $f_\mathrm{PBH}$, can be in principle quite significant: an increase (decrease) of $\csin$ by a factor of $3$ weakens (strengthens) the bound by two orders of magnitude. 

Given the sensitivity of the bounds to the precise choice of $\csin$, additional simulations of the radiative feedback from the accretion onto PBHs adapted to the specific cosmological settings are needed.  

\subsection{Radiative efficiency}
As explained {in \autoref{sec:Einjection}}, the efficiency of the energy transfer to the leptonic component of the accretion flow, usually called $\delta$, is a key parameter. In fact, it controls the radiative efficiency, and ultimately the amount of gravitational binding energy that is not advected into the {PBH} but ultimately radiated and absorbed in the surrounding medium. 

We display in the mid-row of \autoref{fig:panel_figure} the impact of this parameter, $\delta$, on the {constraint on the fractional abundance of PBHs}.
Adopting the three reference values extracted from~\cite{Xie_2012}, the position of the upper limit is shifted by roughly an order of magnitude.
However, given the discussion in \autoref{sec:Einjection} about the physical interpretation of the parameter, it turns out that, despite a preferential transfer of energy to the ions, the electron plasma can also be heated up by a variety of kinetic processes in the context of MHD turbulence. Therefore, the very low values of these parameters that would in principle allow for a significant weakening of the bound {do} not appear to be realistic, and we can conclude that the result is reasonably robust with respect to the uncertainties associated with radiative cooling.

\subsection{Energy deposition function}
The impact of the different methods that we consider for the computation of the energy deposition into the medium is shown in the bottom row of \autoref{fig:panel_figure}. The different recipes described in \autoref{sec:Edeposition} produce result that differ by a factor of a few, and the overall uncertainty on the final result is slightly less than an order of magnitude.

We note that the choice of the {\it extrapolated} transfer functions decreases the value of the energy deposition functions, thus reducing the deposited energy, and weakening the bound. This is due to  $0 < T\left(z^{\prime}, z, \omega\right)  < 1$ in the extrapolated region ($\omega < 5$ keV) of Eq.~\eqref{eq:f_c integral}.

\subsection{A conservative cosmological bound}\label{sec:conservative}
It is important to point out that all the results discussed so far rely on the hypothesis that an {\it accretion disk} is formed.
We {now} relax this assumption, and assess how {the bound weakens} in the case of {\it spherical accretion} within our framework. Given all the previous discussion on the uncertainties, we aim at computing the {\it most conservative upper limit} within the physical setup considered in this work. 
In order to do so, we adopt the following setup. As far as the accretion rate and radiative efficiency are concerned, we refer to the formalism presented in~\cite{AliHaimoudKamionkowski2017} for the case of spherical accretion. That paper presents an analytical computation of both the fudge factor $\lambda$ that appears in the BHL formula and the radiative efficiency $\epsilon$. The recipe is valid {\it in the absence of an accretion disk} and is characterized by a much lower emission of photons with respect to the disk case. The $\lambda$ factor takes into account the presence of the Compton drag and Compton cooling by CMB photons, and the radiative efficiency formula captures the physics of free-free emission, the main process at work within these assumptions. Regarding the energy deposition, we also adopt the most conservative prescription, i.e. the {\it extrapolated transfer function}, which, as shown above, determines the weakest upper limit.

Motivated by the results discussed in \autoref{sec:mini-halos} and \autoref{sec:resultsBounds}, we make the conservative choice to entirely neglect the effect of the DM mini-halos on the accretion rate. 
We show the resulting constraint as a solid dark blue line in \autoref{fig:mainPlot}. In this very conservative setup, the bound is weakened by more than an order of magnitude, but still quite strong, especially for large masses. The most conservative upper limit is still the most stringent available bound towards $M = 10^4 \, \Msun$. Beyond this threshold, our examination of the CMB, treating PBHs as a DM fluid component, would likely break down.

\clearpage

\section{Discussion and conclusions}
\label{sec:discussion}
In this work, we have critically reviewed the ingredients at the core of the cosmological constraint on accreting PBHs. This bound stems from the energy injected by the PBHs into the inter-galactic medium during the dark ages, which affects both the optical depth and the visibility function of CMB photons.
Such a bound depends a priori on a variety of astrophysical details. We discussed the key role of the modeling of baryonic accretion as well as the contribution from DM mini-halos when the fraction of DM that is in the form of PBHs, $\fPBH$, satisfies $\fPBH \ll 1$, inspecting for the first time the interplay between the two, see \autoref{fig:potential} and \autoref{fig:MdotWHalo}.
We have derived the CMB bound on a handful of realistic physics cases, paying attention once again to the interplay of baryonic accretion with the formation of DM mini-halos, see~\autoref{fig:mainPlot}. 

When the effect of DM mini-halos around PBHs is taken into account, \textit{the constraint obtained within the PR model is weaker by up to 3 orders of magnitude than that obtained within the BHL model}.
Moreover, our analysis suggests that the BHL estimate is likely to be overestimating the impact of DM mini-halos, as the local increase of the sound speed has not been taken into account. The PR model provides a simple way to account for this increase. In a model-independent fashion, we can conclude that \textit{the  local increase in temperature and ionization fraction, which we expect to be induced by the accretion process, drastically reduces the impact of DM mini halos on the accretion rate and hence on the CMB bound on PBHs}. 

We have then systematically investigated the impact of what we {believe} to be the key astrophysical ingredients entering in the cosmological analysis of massive PBHs, quantifying their effect on $\fPBH$ for a reference PBH mass of $10^3 M_{\odot}$ in \autoref{fig:panel_figure}. 
Finally, on the basis of our exploration of the main astrophysical phenomena at stake, we singled out a conservative bound on massive PBHs from the CMB, highlighted in dark blue in \autoref{fig:mainPlot}.

It seems natural for us to conclude that despite the multitude of astrophysical uncertainties and the technicalities of the numerical analysis involved, the CMB can easily rule out massive PBHs as light as 10 $M_{\odot}$ as DM candidates. However, it remains true that a very conservative approach to the physics of the bound can robustly exclude $\fPBH > 0.1$ only for PBHs of mass equal or greater than $6 \times 10^{2} M_{\odot}$ at 95\% probability. 

In light of these conclusions, {we note the following}:
\begin{itemize}
\item The relative velocity between DM and baryons plays a crucial role in our treatment. Therefore, one may wonder how to quantify the uncertainty on this parameter. The {first} issue is that the linear approximation adopted in this work inevitably breaks down at late time when virialized halos start to form. {The second issue is that} it may not apply at the small scales relevant for accretion. {These issues} are particularly complex and require dedicated simulations to be properly addressed. The {first issue}  does not seem to impact the result presented here, since the effect on the CMB is dominated by the energy injection that happens at $z > 200$, prior to the formation of large virialized halos with significant velocity {dispersions}. Regarding the {second issue}, the authors of~\cite{Hutsi:2019hlw,Inman:2019wvr} have shown that the presence of PBHs and their clustering may imply an enhancement of the power spectrum at small scales, but it is unclear whether these early-forming structures can acquire a significant baryon content at early times. The results of the simulations seem to suggest that the non-linear small-scale corrections to the relative DM-baryon velocity associated with this effect are subdominant with respect to the {corrections at large scales}, and {thus} the impact on the CMB bound is considered negligible in these works. Further studies are certainly needed in this direction to better clarify and quantify these issues.
\item The (possibly combined) impact of a broad mass function and/or a significant initial clustering on the results presented here might be non-trivial.
The effect of a broader mass function, for instance a log-normal centered around a specific mass value, has been extensively discussed in the literature, and remapping procedures were devised in order to properly remap the bound~\cite{Bellomo:2017zsr}. The upper limits do not show a significant weakening when these procedures are applied. 
It is more complicated to assess whether less trivial mass functions featuring multiple peaks at relevant mass scales possibly associated with modifications of the equation of state at some critical events in the early universe thermal history~\cite{Carr:2019kxo,Niemeyer:1997mt,Franciolini:2022tfm} can evade the bound. In particular, the combination of such peculiar shapes for the mass function and a significant initial clustering of the PBHs may in principle alter all the current upper limits (see discussions on initial clustering and related phenomenology in some specific formation models, e.g. in~\cite{Chisholm:2005vm,Ballesteros:2018swv,Clesse:2016vqa}). 
\end{itemize}
Whether the aforementioned points may effectively change any of the conclusions drawn from the outcome reported in \autoref{fig:mainPlot} poses an interesting question. A convincing answer to that likely requires dedicated N-body simulations, which go beyond the scope of the present study.

\acknowledgments
D.G.~acknowledges support from the project ``Theoretical Astroparticle Physics (TAsP)'' funded by INFN. D.A.~acknowledges support from the Generalitat Valenciana under the grant CIGRIS/2021/054. M.V. acknowledges support from the project ``Theoretical Particle Physics and Cosmology (TPPC)'' funded by INFN. 
R.E.~acknowledges support from the US Department of Energy (DOE) under Grant DE-SC0009854, 
from the Heising-Simons Foundation under Grant No.~79921, from the Simons Foundation under the Simons Investigator in Physics Award~623940, and from the Binational Science Foundation under Grant No.\ 2020220.  G.S.~was supported in part by the DOE under Grant DE-SC0009854 and the Simons Foundation under the Simons Investigator in Physics Award~623940. 
D.G., F.S, and M.V. are grateful to the organizers of the workshop ``Future Perspectives on Primordial Black Holes'' for fruitful and vibrant discussions on PBHs.  F.S acknowledges support from the ANR project GaDaMa (ANR-18-CE31-0006).
We are particularly grateful to the authors of Ref.s~\cite{Facchinetti2022,poulin_cmb_2017} for very fruitful exchange and/or communication during all the stages of the present work.

\clearpage
%\onecolumngrid
\appendix

\section{Details on the Park-Ricotti model}
\label{app:PR_model}
The PR accretion rate is given by 
\begin{equation}
\label{eq:ricotti}
\dot{M}_\mathrm{PR} = 4 \pi \frac {(GM)^2 \rhoin} {(\vin^2 + \csin^2)^{3/2}} \,\,\, ,
\end{equation}
% %
where $\vin, \rhoin$, and $\csin$ are, respectively, the values of the BH velocity, gas density, and sound speed within the ionized region. 
\begin{figure}[ht!]
\centering
\includegraphics[width=.7\linewidth]{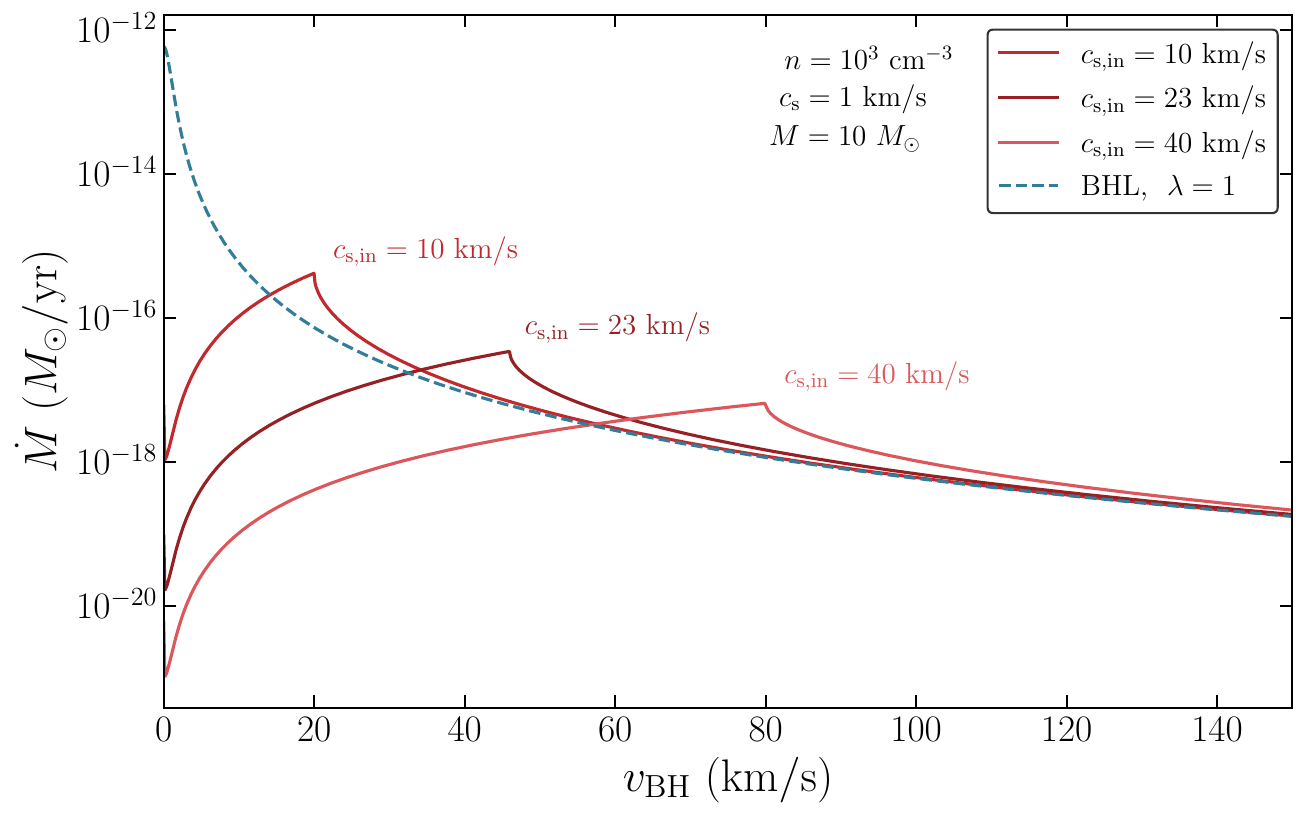}
\caption{The PR accretion rate as a function of the BH-gas relative speed, for different values of the sound speed of the ionized gas. The BHL rate is shown for comparison.}
\label{fig:pr_vbh_cs}
\end{figure}
Three regimes exist (see~\cite{park_accretion_2013, Scarcella:2023voq} for more details), depending on the value of the black hole speed with respect to the gas:    
\begin{itemize}
\item high-velocity regime: $\vrel \geq \vR \approx 2 \csin$
\begin{equation}
\begin{cases}
    \quad
    \rhoin= \rho \dfrac{\vrel^2+\cs^2 - \sqrt{\Delta}}{2 \, \csin^2 }   , \\[5pt]
    \quad  
    \vin= \dfrac{\rho}{\rhoin}{\vrel}  \; 
\end{cases}
\end{equation}
\item low-velocity regime: $\vrel \leq \vD \approx \cs^2/(2 \csin) $
\begin{equation}
\begin{cases}
    \quad
    \rhoin= \rho \dfrac{\vrel^2+\cs^2 + \sqrt{\Delta}}{2 \, \csin^2 } , \\[5pt]
    \quad  
    \vin= \dfrac{\rho}{\rhoin}{\vrel}   \; 
\end{cases}
\end{equation}
\item intermediate velocity regime (a shock front is formed):  $ \vD < \vrel <\vR$
\begin{equation}
\begin{cases}
    \quad
    \rhoin= \rho \dfrac{\vrel^2+\cs^2 }{2 \, \csin^2 }  , \\[5pt]
    \quad  
    \vin= \csin  \; 
\end{cases}
\end{equation}

\end{itemize}
where $\Delta$ is given by
\begin{equation}
\Delta = \sqrt{(\vrel^2 +\cs^2)^2 -4 \vrel^2 \csin^2}
\end{equation}
and $\vR, \vD$ , with $\vR > \vD$ are the values for which $\Delta=0$.

\section{Details on the numerical analysis}\label{app:mcmc_details}
In this work, we made use of the Bayesian code \texttt{Cobaya}~\cite{Torrado2020,2019ascl.soft10019T} for our statistical analysis, adopting the \texttt{CosmoMC} algorithm~\cite{Lewis:2013hha} as the Monte Carlo Markov Chain (MCMC) sampler. We also took advantage of \texttt{getdist}~\cite{Lewis:2019xzd} for processing the resulting chains. Following the Planck collaboration~\cite{planck_collaboration_planck_2020}, we sampled our  $\Lambda$CDM cosmological parameters with a bounded uniform flat prior, as reported in Table~\ref{tab:priors}. 

We used the same $\Lambda$CDM parameters  as the Planck collaboration, with the exception of sampling directly over the Hubble constant $h$ instead of the  angular acoustic scale $\theta_{*}$, and over the redshift of reionization $z_{\rm reio}$ instead of the optical depth to reionization $\tau_{\rm reio}$. This choice avoided a \textit{shooting} problem encountered in \texttt{CLASS} when computing the reionization history.\footnote{See the documentation of \texttt{CLASS} for more details~\cite{blas_cosmic_2011}.}
\begin{table}[t!]
\renewcommand{\arraystretch}{2}
\centering
\label{tab:priors}
\begin{tabular}{|c|c|c|c|c|c|c|}
\hline
\boldmath${\Omega_{b}} \times 10^2$ & \boldmath$\Omega_{\rm c} \times 10^2$ & \boldmath${h}$  &  \boldmath$\ln (10^{10}A_{\rm s})$ & \boldmath$n_{s}$  & \boldmath$z_\mathrm{reio} $ & \boldmath$f_\mathrm{PBH} $  \\
\hline
\hline
$\mathcal{U}(0.5,4)$ & $\mathcal{U}(0.1,36)$ & $\mathcal{U}(0.6,0.8)$ & $\mathcal{U}(1.61,3.91)$ & $\mathcal{U}(0.8,1.2)$ & $\mathcal{U}(4,10)$ & $\mathcal{U}(0,1)$ \\
\hline
\end{tabular}
\caption{Cosmological parameters varied in this work and their respective priors. $\mathcal{U}$ denotes a uniform distribution between the specified \texttt{min} and \texttt{max} values.}
\label{tab:priors}
\end{table}
 We verified that $\Lambda$CDM parameters are mostly unchanged when PBHs are included, and when using different accretion models, as shown in  \autoref{fig:trianglePlot}. With this in mind, we provided covariance matrices of the proposal posterior distribution functions (pdfs) for all known cosmological and nuisance parameters, which we derived from running a long $\Lambda$CDM run using all our cosmological likelihoods (see the list and references in \autoref{sec:CLASS}). This method allows for more efficient sampling, when extending to the 7-dimensional  $\Lambda$CDM + $\fPBH$ parameter space, and we manually entered an estimate for the reference and proposal for $\fPBH$.\footnote{The value of the proposal defines a proposal pdf which is a Gaussian mixed with an exponential in random directions, which is less sensitive to poor initial-width estimates~\cite{Neal2005,Lewis2002,Lewis2013}.} 
\begin{figure}[h!]
\centering
\includegraphics[width=\linewidth]{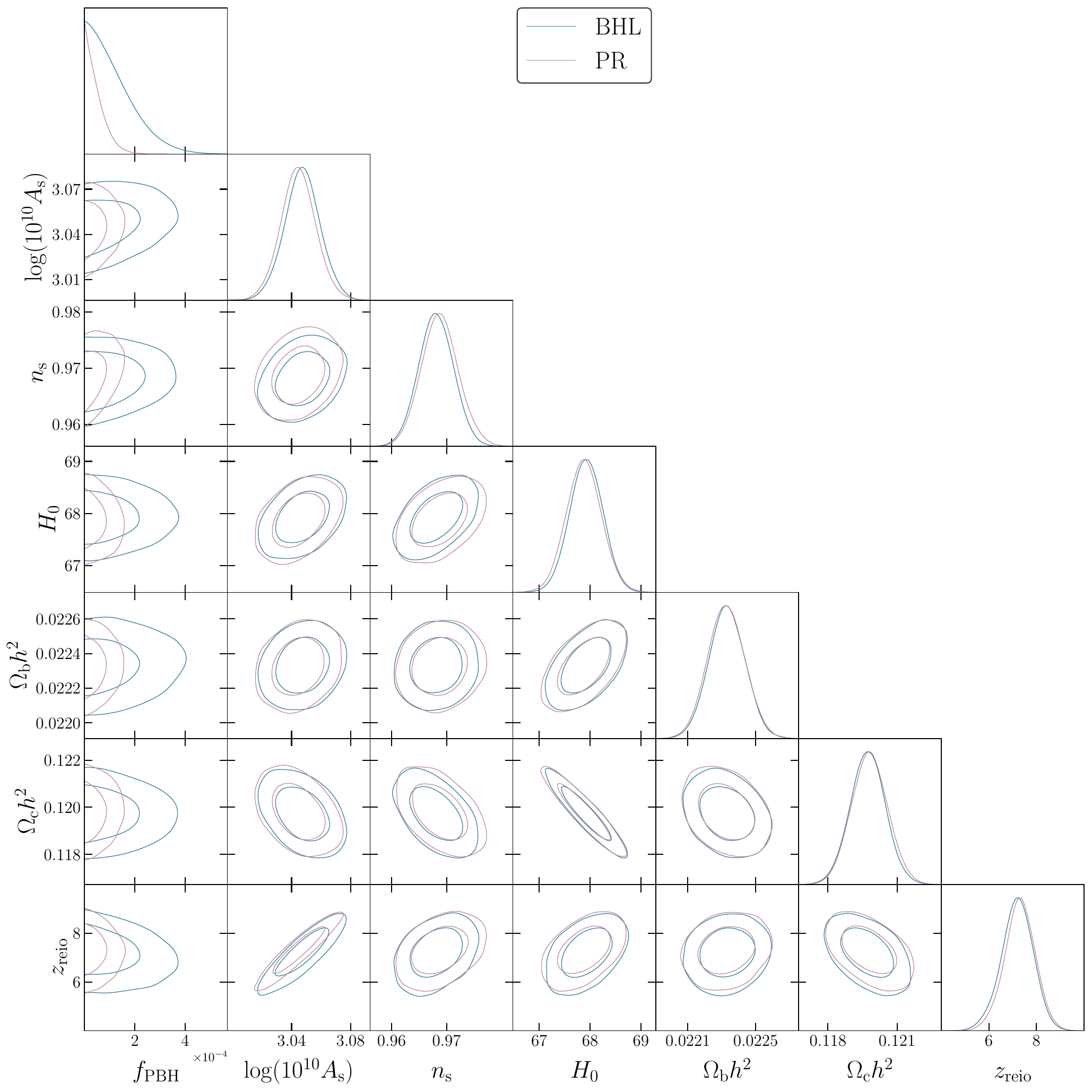}
\caption{ A triangle plot comparing the posteriors at 68\% and 95\% at $M = 10^3  \ M_\odot$ for BHL and PR. }
\label{fig:trianglePlot}
\end{figure}

\section{Comparison to previous works}\label{app:comparison}
Considerable effort was made to conduct a consistency check with previous works~\cite{poulin_cmb_2017,Facchinetti2022}. In this appendix, we compare our benchmark model to those presented in~\cite{Facchinetti2022, poulin_cmb_2017}, and we comment on our attempt to replicate their results exactly. As shown in \autoref{fig:appendix_bound} our benchmark results are within approximately an order of magnitude to those in the literature. There are, however, many important differences that distinguish our recipes. Roughly speaking, discrepancies can be attributed  to our different treatments of the energy deposition functions, PBH relative velocities, recombination code and \texttt{CLASS} versions, and our use of more recent and different likelihoods. We provide a more detailed description of these discrepancies below. 

\subsection{Comparison to Ref.~\cite{Facchinetti2022}} 
In this section, FLC2023 refers to the PR model implemented in~\cite{Facchinetti2022}. Although the underlying PR accretion rate implemented in our work does not differ from FLC2023,\footnote{We use the same sound speed in the ionized region $\csin = 23 \ {\rm km/s}$.  } there is an important difference in the technique used to average over the PBH velocity distribution. As previously shown in~\cite{AliHaimoudKamionkowski2017}, incorporating PBH velocities in the Bondi formula is done through the transformation,  
\begin{equation}
 \dot{M}_{\rm B} \sim \frac{1}{c_{\rm s }^3} \quad \quad \to \quad \quad \dot{M}_{\rm BHL} \sim \frac{1}{(c_{\rm s}^2 + v_{\rm PBH}^2)^{3/2}}\ .
\end{equation}
Since the speed of sound is given by Eq. \eqref{eq:soundspeed}, the above transformation is equivalent to 
\begin{equation}
 \dot{M}_{\rm B} \sim \frac{1}{c_{\rm s }^3\left(X_e,T\right)}\quad \quad \to\quad \quad \dot{M}_{\rm BHL} \sim \frac{1}{c_{\rm s}^3\left(X_{e},T+\frac{m_p v_{\rm PBH}^2}{\gamma\left(1+{X}_e\right)}\right)} \ .
\end{equation}
In effect, PBH velocities can be accounted for simply by averaging over the speed of sound in the Bondi model. FLC2023 sets the explicit velocity dependence in the PR model to the root-mean-square (rms) value and then performs this transformation to the speed of sound, integrating it over a Maxwell-Boltzmann distribution.

However, this temperature transformation is only applicable to the Bondi case. In the PR model, the velocity dependence is `built-in', so we simply average over a Maxwell-Boltzmann distribution centered at the baryon-DM relative speed given in Eq. \eqref{eq:vrel}. 

Additionally, FLC2023 uses a radiative efficiency model as described by
\begin{equation}
\epsilon=\epsilon_0 \min \left(1, \frac{\dot{M}_{\mathrm{PBH}}}{L_{\mathrm{Edd}}}\right) 
\label{eq:adaf_facc} ,
\end{equation}
where $\epsilon_0 $ is a scale factor that they fix to $0.1$. In particular, this different functional form means that  the luminosity scales as $L \propto \dot{M}^2 \propto M_{\rm PBH}^4 $ below the Eddington limit. This is contrary to our use of the ADAF formalism as presented in \autoref{sec:Einjection}, where $L \propto \dot{M}^{1+ a} \propto \dot{M}_{\rm PBH}^{2\left(1+ a\right)} $, where $a$ takes a piecewise form depending on the accretion rate (but primarily takes the value $a = 0.59$). This different scaling of the luminosity with $M$, is one of the factors leading to a different slope in \autoref{fig:appendix_bound}. Furthermore,~\cite{Facchinetti2022} uses a different energy deposition treatment, where (referring to the second equality in Eq. \eqref{eqn:dep_fn_defn}) they choose to use the $f_{\rm eff} (z)$ function derived analytically by~\cite{AliHaimoudKamionkowski2017},  with the repartition functions $\chi_c$ computed by~\cite{Galli2013}. There are also differences in the likelihoods, the sampling technique,\footnote{FLC2023 uses a log-spaced prior on $\fPBH$, compared to our uniform linear prior.} and the statistical treatment of the posterior in the computation of the bound. 

In addition to the comparison shown in \autoref{fig:appendix_bound}, we attempted to directly reproduce the results of~\cite{Facchinetti2022}. We thank~\cite{Facchinetti2022} for sharing their version of \texttt{CLASS} with us. For the MCMC sampler,~\cite{Facchinetti2022} used \texttt{MontePython}~\cite{Audren:2012wb, Brinckmann:2018cvx}  to  infer their cosmological parameters, but we checked that the result remains unchanged irrespective of the sampler by comparing directly between the results of \texttt{Cobaya} and \texttt{MontePython}. Using  their exact version of \texttt{CLASS}, as well as a version we modified to replicate their results, we  consistently found a result differing by roughly a factor of $ 4$ across the entire mass range. We attribute this difference to their log sampling and a different statistical treatment of their posterior. 

\subsection{Comparison to Ref.~\cite{poulin_cmb_2017}}
In this section, PSCCK2017 refers to the BHL disk accretion model implemented in~\cite{poulin_cmb_2017}. Despite using the same BHL model and disk accretion geometry, there are a number of important differences that cause our bounds to differ from their bounds. 

The primary reason for this discrepancy is the treatment of PBH velocities. PSCCK2017 uses an approximated effective velocity formula,
\begin{equation}
\dot{M}_{\rm BHL} \sim \frac{1}{v_{\rm eff}^{3}} \quad \quad \text{where}\quad \quad v_{\mathrm{eff}} \equiv\left\langle\frac{1}{\left(c_{\rm s}^2+v_{\mathrm{rel}}^2\right)^3}\right\rangle^{-1 / 6} \simeq \sqrt{c_{\rm s} \sqrt{\left\langle v_{\mathrm{rel}}^2\right\rangle}}\ .
\end{equation} 
This formula implicitly assumes that the total luminosity is quadratic with the accretion rate. Although this is true for a linear efficiency, in the ADAF efficiency model, the scaling is not quadratic and a proper averaging over a Maxwell-Boltzmann distribution is needed.\footnote{This is expanded on in detail in~\cite{Mena_2019}, where the nuance of this calculation is described in their  Eq.~(3.6).} Additionally, the constraint in PSCCK2017 was calculated using different energy deposition functions, as described in \autoref{sec:Edeposition}. In our approach, we cut the transfer functions at 5~keV whereas PSCCK2017 extrapolates down to 100~eV. Moreover, the functional form of the luminosity spectrum used by PSCCK2017 could not be exactly determined. Consequently, we were unable to replicate Fig.~2 in~\cite{poulin_cmb_2017}. Finally, these results relied on Planck 2015 likelihoods, whereas we use the more recent Planck 2018 likelihoods as well as other cosmological data sets. Considerable efforts were made to compare directly to this work by implementing to the best of our ability the recipe used by~\cite{poulin_cmb_2017}. Unfortunately, an exact replication of their results was not possible due to some uncertainties regarding their recipe, but for our best attempt, our bounds differed only by an overall $\mathcal{O}(1)$ factor.

\begin{figure}[t!]
\centering
\includegraphics[width=\linewidth]{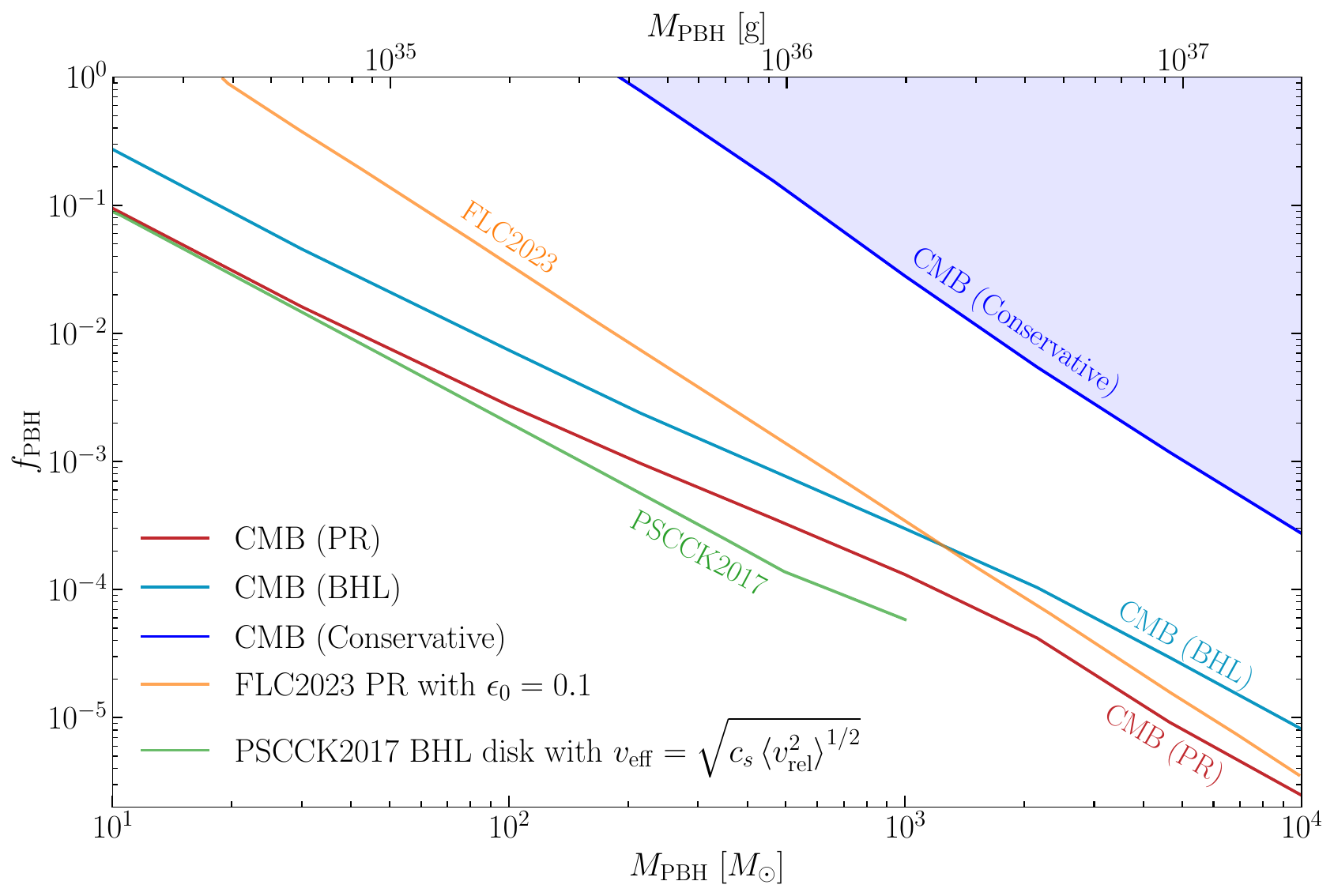}
\caption{Bounds at 95\% probability on the abundance of a monochromatic distribution of PBHs, with a comparison to previous works. In light blue, we show our bound assuming BHL accretion with disk geometry without DM mini-halos, in red, we show the bound assuming PR accretion with disk geometry without DM mini-halos, and in dark blue, we show the conservative bound assuming spherical accretion [equivalent to those shown in \autoref{fig:mainPlot}]. In orange, we show the bound computed by~\cite{Facchinetti2022} (FLC2023) using the PR model and reference value of $\epsilon_0 = 0.1$. In green, we show the bound computed by~\cite{poulin_cmb_2017} (PSCCK2017), using the BHL model with disk accretion, and $v_{\rm eff} = \sqrt{c_{s, \infty}\left\langle v_L^2\right\rangle^{1 / 2}}$. There are a number of differences between these recipes, which we describe in detail in \autoref{app:comparison}.}
\label{fig:appendix_bound}
\end{figure}

\bibliographystyle{JHEP}
\bibliography{lib.bib}

\providecommand{\href}[2]{#2}\begingroup\raggedright\begin{thebibliography}{100}

\bibitem{Facchinetti2022}
G.~Facchinetti, M.~Lucca and S.~Clesse, \emph{Relaxing cmb bounds on primordial black holes: the role of ionization fronts}, \href{http://dx.doi.org/10.1103/physrevd.107.043537}{\emph{Physical Review D} {\bfseries 107} (Feb., 2022) 043537}, [\href{https://arxiv.org/abs/2212.07969}{{\ttfamily 2212.07969}}].

\bibitem{poulin_cmb_2017}
V.~Poulin, P.~D. Serpico, F.~Calore, S.~Clesse and K.~Kohri, \emph{{CMB} bounds on disk-accreting massive {Primordial} {Black} {Holes}}, \href{http://dx.doi.org/10.1103/PhysRevD.96.083524}{\emph{Physical Review D} {\bfseries 96} (Oct., 2017) 083524}.

\bibitem{Peebles:2000pn}
P.~J.~E. Peebles, S.~Seager and W.~Hu, \emph{{Delayed recombination}}, \href{http://dx.doi.org/10.1086/312831}{\emph{Astrophys. J. Lett.} {\bfseries 539} (2000) L1--L4}, [\href{https://arxiv.org/abs/astro-ph/0004389}{{\ttfamily astro-ph/0004389}}].

\bibitem{Pierpaoli_2004}
E.~Pierpaoli, \emph{Decaying particles and the reionization history of the universe}, \href{http://dx.doi.org/10.1103/physrevlett.92.031301}{\emph{Physical Review Letters} {\bfseries 92} (Jan., 2004) 031301}.

\bibitem{Chen:2003gz}
X.-L. Chen and M.~Kamionkowski, \emph{{Particle decays during the cosmic dark ages}}, \href{http://dx.doi.org/10.1103/PhysRevD.70.043502}{\emph{Phys. Rev. D} {\bfseries 70} (2004) 043502}, [\href{https://arxiv.org/abs/astro-ph/0310473}{{\ttfamily astro-ph/0310473}}].

\bibitem{Padmanabhan_2005}
N.~Padmanabhan and D.~P. Finkbeiner, \emph{Detecting dark matter annihilation with cmb polarization: Signatures and experimental prospects}, \href{http://dx.doi.org/10.1103/physrevd.72.023508}{\emph{Physical Review D} {\bfseries 72} (July, 2005) 023508}.

\bibitem{Zhang_2006}
L.~Zhang, X.~Chen, Y.-A. Lei and Z.-g. Si, \emph{Impacts of dark matter particle annihilation on recombination and the anisotropies of the cosmic microwave background}, \href{http://dx.doi.org/10.1103/physrevd.74.103519}{\emph{Physical Review D} {\bfseries 74} (Nov., 2006) 103519}.

\bibitem{Zhang_2007}
L.~Zhang, X.~Chen, M.~Kamionkowski, Z.-g. Si and Z.~Zheng, \emph{Constraints on radiative dark-matter decay from the cosmic microwave background}, \href{http://dx.doi.org/10.1103/physrevd.76.061301}{\emph{Physical Review D} {\bfseries 76} (Sept., 2007) 061301}.

\bibitem{1981MNRAS.194..639C}
B.~J. {Carr}, \emph{{Pregalactic black hole accretion and the thermal history of the universe}}, \href{http://dx.doi.org/10.1093/mnras/194.3.639}{\emph{\mnras} {\bfseries 194} (Feb., 1981) 639--668}.

\bibitem{Ricotti:2007jk}
M.~Ricotti, \emph{{Bondi accretion in the early universe}}, \href{http://dx.doi.org/10.1086/516562}{\emph{Astrophys. J.} {\bfseries 662} (2007) 53--61}, [\href{https://arxiv.org/abs/0706.0864}{{\ttfamily 0706.0864}}].

\bibitem{Chen:2016pud}
L.~Chen, Q.-G. Huang and K.~Wang, \emph{{Constraint on the abundance of primordial black holes in dark matter from Planck data}}, \href{http://dx.doi.org/10.1088/1475-7516/2016/12/044}{\emph{JCAP} {\bfseries 12} (2016) 044}, [\href{https://arxiv.org/abs/1608.02174}{{\ttfamily 1608.02174}}].

\bibitem{Horowitz:2016lib}
B.~Horowitz, \emph{{Revisiting Primordial Black Holes Constraints from Ionization History}},  \href{https://arxiv.org/abs/1612.07264}{{\ttfamily 1612.07264}}.

\bibitem{AliHaimoudKamionkowski2017}
Y.~Ali-Haïmoud and M.~Kamionkowski, \emph{Cosmic microwave background limits on accreting primordial black holes}, \href{http://dx.doi.org/10.1103/physrevd.95.043534}{\emph{Physical Review D} {\bfseries 95} (feb, 2017) }.

\bibitem{Serpico:2020ehh}
P.~D. Serpico, V.~Poulin, D.~Inman and K.~Kohri, \emph{{Cosmic microwave background bounds on primordial black holes including dark matter halo accretion}}, \href{http://dx.doi.org/10.1103/PhysRevResearch.2.023204}{\emph{Phys. Rev. Res.} {\bfseries 2} (2020) 023204}, [\href{https://arxiv.org/abs/2002.10771}{{\ttfamily 2002.10771}}].

\bibitem{Hoyle1939}
F.~Hoyle and R.~A. Lyttleton, \emph{The effect of interstellar matter on climatic variation}, \href{http://dx.doi.org/10.1017/s0305004100021150}{\emph{Mathematical Proceedings of the Cambridge Philosophical Society} {\bfseries 35} (July, 1939) 405--415}.

\bibitem{1952MNRAS.112..195B}
H.~{Bondi}, \emph{{On spherically symmetrical accretion}}, \href{http://dx.doi.org/10.1093/mnras/112.2.195}{\emph{\mnras} {\bfseries 112} (Jan., 1952) 195}.

\bibitem{bondi_spherically_1952}
H.~Bondi, \emph{On spherically symmetrical accretion}, \href{http://dx.doi.org/10.1093/mnras/112.2.195}{\emph{Monthly Notices of the Royal Astronomical Society} {\bfseries 112} (Jan., 1952) 195}.

\bibitem{park_accretion_2011}
K.~Park and M.~Ricotti, \emph{Accretion onto {Black} {Holes} from {Large} {Scales} {Regulated} by {Radiative} {Feedback}. {I}. {Parametric} {Study} for {Spherically} {Symmetric} {Accretion}}, \href{http://dx.doi.org/10.1088/0004-637X/739/1/2}{\emph{The Astrophysical Journal} {\bfseries 739} (Aug., 2011) 2}.

\bibitem{park_accretion_2012}
K.~Park and M.~Ricotti, \emph{Accretion onto {Black} {Holes} from {Large} {Scales} {Regulated} by {Radiative} {Feedback}. {II}. {Growth} {Rate} and {Duty} {Cycle}}, \href{http://dx.doi.org/10.1088/0004-637X/747/1/9}{\emph{The Astrophysical Journal} {\bfseries 747} (Mar., 2012) 9}.

\bibitem{park_accretion_2013}
K.~Park and M.~Ricotti, \emph{Accretion onto {Black} {Holes} from {Large} {Scales} {Regulated} by {Radiative} {Feedback}. {III}. {Enhanced} {Luminosity} of {Intermediate} {Mass} {Black} {Holes} {Moving} at {Supersonic} {Speeds}}, \href{http://dx.doi.org/10.1088/0004-637X/767/2/163}{\emph{The Astrophysical Journal} {\bfseries 767} (Apr., 2013) 163}.

\bibitem{Mack:2006gz}
K.~J. Mack, J.~P. Ostriker and M.~Ricotti, \emph{{Growth of structure seeded by primordial black holes}}, \href{http://dx.doi.org/10.1086/518998}{\emph{Astrophys. J.} {\bfseries 665} (2007) 1277--1287}, [\href{https://arxiv.org/abs/astro-ph/0608642}{{\ttfamily astro-ph/0608642}}].

\bibitem{Xie_2012}
F.-G. Xie and F.~Yuan, \emph{Radiative efficiency of hot accretion flows}, \href{http://dx.doi.org/10.1111/j.1365-2966.2012.22030.x}{\emph{Monthly Notices of the Royal Astronomical Society} {\bfseries 427} (nov, 2012) 1580--1586}.

\bibitem{Slatyer:2012yq}
T.~R. Slatyer, \emph{{Energy Injection And Absorption In The Cosmic Dark Ages}}, \href{http://dx.doi.org/10.1103/PhysRevD.87.123513}{\emph{Phys. Rev. D} {\bfseries 87} (2013) 123513}, [\href{https://arxiv.org/abs/1211.0283}{{\ttfamily 1211.0283}}].

\bibitem{slatyer_indirect_2016}
T.~R. Slatyer, \emph{Indirect {Dark} {Matter} {Signatures} in the {Cosmic} {Dark} {Ages} {II}. {Ionization}, {Heating} and {Photon} {Production} from {Arbitrary} {Energy} {Injections}}, \href{http://dx.doi.org/10.1103/PhysRevD.93.023521}{\emph{Physical Review D} {\bfseries 93} (Jan., 2016) 023521}.

\bibitem{Finkbeiner_2012}
D.~P. Finkbeiner, S.~Galli, T.~Lin and T.~R. Slatyer, \emph{Searching for dark matter in the cmb: A compact parametrization of energy injection from new physics}, \href{http://dx.doi.org/10.1103/physrevd.85.043522}{\emph{Physical Review D} {\bfseries 85} (Feb., 2012) }.

\bibitem{Bellomo:2017zsr}
N.~Bellomo, J.~L. Bernal, A.~Raccanelli and L.~Verde, \emph{Primordial black holes as dark matter: converting constraints from monochromatic to extended mass distributions}, \href{http://dx.doi.org/10.1088/1475-7516/2018/01/004}{\emph{Journal of Cosmology and Astroparticle Physics} {\bfseries 2018} (Jan., 2018) 004–004}.

\bibitem{K_hnel_2017}
F.~Kühnel and K.~Freese, \emph{Constraints on primordial black holes with extended mass functions}, \href{http://dx.doi.org/10.1103/physrevd.95.083508}{\emph{Physical Review D} {\bfseries 95} (Apr., 2017) }.

\bibitem{Carr_2017}
B.~Carr, M.~Raidal, T.~Tenkanen, V.~Vaskonen and H.~Veermäe, \emph{Primordial black hole constraints for extended mass functions}, \href{http://dx.doi.org/10.1103/physrevd.96.023514}{\emph{Physical Review D} {\bfseries 96} (July, 2017) }.

\bibitem{Tseliakhovich2010}
D.~Tseliakhovich and C.~Hirata, \emph{Relative velocity of dark matter and baryonic fluids and the formation of the first structures}, \href{http://dx.doi.org/10.1103/physrevd.82.083520}{\emph{Phys.Rev.D82:083520,2010} {\bfseries 82} (Oct., 2010) 083520}, [\href{https://arxiv.org/abs/1005.2416}{{\ttfamily 1005.2416}}].

\bibitem{Dvorkin2013}
C.~Dvorkin, K.~Blum and M.~Kamionkowski, \emph{Constraining dark matter-baryon scattering with linear cosmology}, \href{http://dx.doi.org/10.1103/physrevd.89.023519}{\emph{Phys.Rev.D89:023519,2014} {\bfseries 89} (Jan., 2013) 023519}, [\href{https://arxiv.org/abs/1311.2937}{{\ttfamily 1311.2937}}].

\bibitem{Perna_2003}
R.~Perna, R.~Narayan, G.~Rybicki, L.~Stella and A.~Treves, \emph{Bondi accretion and the problem of the missing isolated neutron stars}, \href{http://dx.doi.org/10.1086/377091}{\emph{The Astrophysical Journal} {\bfseries 594} (sep, 2003) 936}.

\bibitem{Fender:2013ei}
R.~Fender, T.~Maccarone and I.~Heywood, \emph{{The closest black holes}}, \href{http://dx.doi.org/10.1093/mnras/sts688}{\emph{Mon. Not. Roy. Astron. Soc.} {\bfseries 430} (2013) 1538}, [\href{https://arxiv.org/abs/1301.1341}{{\ttfamily 1301.1341}}].

\bibitem{Pellegrini_2005}
S.~Pellegrini, \emph{Nuclear accretion in galaxies of the local universe: Clues from chandra observations}, \href{http://dx.doi.org/10.1086/429267}{\emph{The Astrophysical Journal} {\bfseries 624} (may, 2005) 155}.

\bibitem{Wang:2013dqq}
Q.~D. Wang et~al., \emph{{Dissecting X-ray-emitting Gas around the Center of our Galaxy}}, \href{http://dx.doi.org/10.1126/science.1240755}{\emph{Science} {\bfseries 341} (2013) 981}, [\href{https://arxiv.org/abs/1307.5845}{{\ttfamily 1307.5845}}].

\bibitem{RicottiOstrikerMack2008ApJ}
M.~{Ricotti}, J.~P. {Ostriker} and K.~J. {Mack}, \emph{{Effect of Primordial Black Holes on the Cosmic Microwave Background and Cosmological Parameter Estimates}}, \href{http://dx.doi.org/10.1086/587831}{\emph{Astrophys. J.} {\bfseries 680} (June, 2008) 829--845}, [\href{https://arxiv.org/abs/0709.0524}{{\ttfamily 0709.0524}}].

\bibitem{park_thesis_2012}
K.~Park, \emph{Accretion onto Black Holes from Large Scales Regulated by Radiative Feedback}.
\newblock Phd thesis, University of Maryland, 2012.

\bibitem{Berezinsky_2013}
V.~Berezinsky, V.~Dokuchaev and Y.~Eroshenko, \emph{Formation and internal structure of superdense dark matter clumps and ultracompact minihaloes}, \href{http://dx.doi.org/10.1088/1475-7516/2013/11/059}{\emph{Journal of Cosmology and Astroparticle Physics} {\bfseries 2013} (Nov., 2013) 059–059}.

\bibitem{Delos:2017thv}
M.~S. Delos, A.~L. Erickcek, A.~P. Bailey and M.~A. Alvarez, \emph{{Are ultracompact minihalos really ultracompact?}}, \href{http://dx.doi.org/10.1103/PhysRevD.97.041303}{\emph{Phys. Rev. D} {\bfseries 97} (2018) 041303}, [\href{https://arxiv.org/abs/1712.05421}{{\ttfamily 1712.05421}}].

\bibitem{Adamek:2019gns}
J.~Adamek, C.~T. Byrnes, M.~Gosenca and S.~Hotchkiss, \emph{{WIMPs and stellar-mass primordial black holes are incompatible}},  \href{https://arxiv.org/abs/1901.08528}{{\ttfamily 1901.08528}}.

\bibitem{Boudaud:2021irr}
M.~Boudaud, T.~Lacroix, M.~Stref, J.~Lavalle and P.~Salati, \emph{{In-depth analysis of the clustering of dark matter particles around primordial black holes. Part~I. Density profiles}}, \href{http://dx.doi.org/10.1088/1475-7516/2021/08/053}{\emph{JCAP} {\bfseries 08} (2021) 053}, [\href{https://arxiv.org/abs/2106.07480}{{\ttfamily 2106.07480}}].

\bibitem{2019PhRvD.100b3506A}
J.~{Adamek}, C.~T. {Byrnes}, M.~{Gosenca} and S.~{Hotchkiss}, \emph{{WIMPs and stellar-mass primordial black holes are incompatible}}, \href{http://dx.doi.org/10.1103/PhysRevD.100.023506}{\emph{\prd} {\bfseries 100} (July, 2019) 023506}, [\href{https://arxiv.org/abs/1901.08528}{{\ttfamily 1901.08528}}].

\bibitem{1985ApJS...58...39B}
E.~{Bertschinger}, \emph{{Self-similar secondary infall and accretion in an Einstein-de Sitter universe}}, \href{http://dx.doi.org/10.1086/191028}{\emph{\apjs} {\bfseries 58} (May, 1985) 39--65}.

\bibitem{Carr_2021}
B.~Carr, F.~Kühnel and L.~Visinelli, \emph{Black holes and wimps: all or nothing or something else}, \href{http://dx.doi.org/10.1093/mnras/stab1930}{\emph{Monthly Notices of the Royal Astronomical Society} {\bfseries 506} (July, 2021) 3648–3661}.

\bibitem{Eroshenko_2016}
Y.~N. Eroshenko, \emph{Dark matter density spikes around primordial black holes}, \href{http://dx.doi.org/10.1134/s1063773716060013}{\emph{Astronomy Letters} {\bfseries 42} (June, 2016) 347–356}.

\bibitem{Boudaud_2021}
M.~Boudaud, T.~Lacroix, M.~Stref, J.~Lavalle and P.~Salati, \emph{In-depth analysis of the clustering of dark matter particles around primordial black holes. part i. density profiles}, \href{http://dx.doi.org/10.1088/1475-7516/2021/08/053}{\emph{Journal of Cosmology and Astroparticle Physics} {\bfseries 2021} (Aug., 2021) 053}.

\bibitem{2016ApJ...818..184P}
K.~{Park}, M.~{Ricotti}, P.~{Natarajan}, T.~{Bogdanovi{\'c}} and J.~H. {Wise}, \emph{{Bulge-driven Fueling of Seed Black Holes}}, \href{http://dx.doi.org/10.3847/0004-637X/818/2/184}{\emph{\apj} {\bfseries 818} (Feb., 2016) 184}, [\href{https://arxiv.org/abs/1512.03434}{{\ttfamily 1512.03434}}].

\bibitem{2020MNRAS.495.2966S}
K.~{Sugimura} and M.~{Ricotti}, \emph{{Structure and instability of the ionization fronts around moving black holes}}, \href{http://dx.doi.org/10.1093/mnras/staa1394}{\emph{\mnras} {\bfseries 495} (July, 2020) 2966--2978}, [\href{https://arxiv.org/abs/2003.05625}{{\ttfamily 2003.05625}}].

\bibitem{1976ApJ...204..555S}
S.~L. {Shapiro} and A.~P. {Lightman}, \emph{{Black holes in X-ray binaries: marginal existence and rotation reversals of accretion disks.}}, \href{http://dx.doi.org/10.1086/154203}{\emph{\apj} {\bfseries 204} (Mar., 1976) 555--560}.

\bibitem{1973A&A....24..337S}
N.~I. {Shakura} and R.~A. {Sunyaev}, \emph{{Black holes in binary systems. Observational appearance.}}, {\emph{\aap} {\bfseries 24} (Jan., 1973) 337--355}.

\bibitem{Arzamasskiy_2019}
L.~Arzamasskiy, M.~W. Kunz, B.~D.~G. Chandran and E.~Quataert, \emph{Hybrid-kinetic simulations of ion heating in alfvénic turbulence}, \href{http://dx.doi.org/10.3847/1538-4357/ab20cc}{\emph{The Astrophysical Journal} {\bfseries 879} (jul, 2019) 53}.

\bibitem{2021ApJ...916..120C}
S.~S. {Cerri}, L.~{Arzamasskiy} and M.~W. {Kunz}, \emph{{On Stochastic Heating and Its Phase-space Signatures in Low-beta Kinetic Turbulence}}, \href{http://dx.doi.org/10.3847/1538-4357/abfbde}{\emph{\apj} {\bfseries 916} (Aug., 2021) 120}, [\href{https://arxiv.org/abs/2102.09654}{{\ttfamily 2102.09654}}].

\bibitem{2019PhRvL.122e5101Z}
V.~{Zhdankin}, D.~A. {Uzdensky}, G.~R. {Werner} and M.~C. {Begelman}, \emph{{Electron and Ion Energization in Relativistic Plasma Turbulence}}, \href{http://dx.doi.org/10.1103/PhysRevLett.122.055101}{\emph{\prl} {\bfseries 122} (Feb., 2019) 055101}, [\href{https://arxiv.org/abs/1809.01966}{{\ttfamily 1809.01966}}].

\bibitem{Galli2013}
S.~Galli, T.~R. Slatyer, M.~Valdes and F.~Iocco, \emph{Systematic uncertainties in constraining dark matter annihilation from the cosmic microwave background}, \href{http://dx.doi.org/10.1103/physrevd.88.063502}{\emph{Physical Review D} {\bfseries 88} (Sept., 2013) 063502}, [\href{https://arxiv.org/abs/1306.0563}{{\ttfamily 1306.0563}}].

\bibitem{stocker_exotic_2018}
P.~Stöcker, M.~Krämer, J.~Lesgourgues and V.~Poulin, \emph{Exotic energy injection with {ExoCLASS}: {Application} to the {Higgs} portal model and evaporating black holes}, \href{http://dx.doi.org/10.1088/1475-7516/2018/03/018}{\emph{Journal of Cosmology and Astroparticle Physics} {\bfseries 2018} (Mar., 2018) 018--018}.

\bibitem{Liu2019}
H.~Liu, G.~W. Ridgway and T.~R. Slatyer, \emph{Darkhistory: A code package for calculating modified cosmic ionization and thermal histories with dark matter and other exotic energy injections}, \href{http://dx.doi.org/10.1103/physrevd.101.023530}{\emph{Phys. Rev. D 101, 023530 (2020)} {\bfseries 101} (Jan., 2019) 023530}, [\href{https://arxiv.org/abs/1904.09296}{{\ttfamily 1904.09296}}].

\bibitem{Capozzi2023}
F.~Capozzi, R.~Z. Ferreira, L.~Lopez-Honorez and O.~Mena, \emph{Cmb and lyman-$\alpha$ constraints on dark matter decays to photons}, \href{http://dx.doi.org/10.1088/1475-7516/2023/06/060}{\emph{Journal of Cosmology and Astroparticle Physics} {\bfseries 2023} (June, 2023) 060}, [\href{https://arxiv.org/abs/2303.07426}{{\ttfamily 2303.07426}}].

\bibitem{Hu:1995kot}
W.~Hu, N.~Sugiyama and J.~Silk, \emph{{The Physics of microwave background anisotropies}}, \href{http://dx.doi.org/10.1038/386037a0}{\emph{Nature} {\bfseries 386} (1997) 37--43}, [\href{https://arxiv.org/abs/astro-ph/9604166}{{\ttfamily astro-ph/9604166}}].

\bibitem{blas_cosmic_2011}
D.~Blas, J.~Lesgourgues and T.~Tram, \emph{The {Cosmic} {Linear} {Anisotropy} {Solving} {System} ({CLASS}) {II}: {Approximation} schemes}, \href{http://dx.doi.org/10.1088/1475-7516/2011/07/034}{\emph{Journal of Cosmology and Astroparticle Physics} {\bfseries 2011} (July, 2011) 034--034}.

\bibitem{Torrado2020}
J.~Torrado and A.~Lewis, \emph{Cobaya: Code for bayesian analysis of hierarchical physical models}, \href{http://dx.doi.org/10.1088/1475-7516/2021/05/057}{\emph{JCAP 05 (2021) 057} {\bfseries 2021} (May, 2020) 057}, [\href{https://arxiv.org/abs/2005.05290}{{\ttfamily 2005.05290}}].

\bibitem{2019ascl.soft10019T}
J.~{Torrado} and A.~{Lewis}, ``{Cobaya: Bayesian analysis in cosmology}.'' Astrophysics Source Code Library, record ascl:1910.019, Oct., 2019.

\bibitem{planck_collaboration_planck_2020}
P.~Collaboration, \emph{Planck 2018 results. {VI}. {Cosmological} parameters}, \href{http://dx.doi.org/10.1051/0004-6361/201833910}{\emph{Astronomy \& Astrophysics} {\bfseries 641} (Sept., 2020) A6}.

\bibitem{beutler_6df_2011}
F.~Beutler, C.~Blake, M.~Colless, D.~H. Jones, L.~Staveley-Smith, L.~Campbell et~al., \emph{The {6dF} {Galaxy} {Survey}: {Baryon} {Acoustic} {Oscillations} and the {Local} {Hubble} {Constant}}, \href{http://dx.doi.org/10.1111/j.1365-2966.2011.19250.x}{\emph{Monthly Notices of the Royal Astronomical Society} {\bfseries 416} (Oct., 2011) 3017--3032}.

\bibitem{ross_clustering_2015}
A.~J. Ross, L.~Samushia, C.~Howlett, W.~J. Percival, A.~Burden and M.~Manera, \emph{The {Clustering} of the {SDSS} {DR7} {Main} {Galaxy} {Sample} {I}: {A} 4 per cent {Distance} {Measure} at z=0.15}, \href{http://dx.doi.org/10.1093/mnras/stv154}{\emph{Monthly Notices of the Royal Astronomical Society} {\bfseries 449} (May, 2015) 835--847}.

\bibitem{alam_clustering_2017}
S.~Alam, M.~Ata, S.~Bailey, F.~Beutler, D.~Bizyaev, J.~A. Blazek et~al., \emph{The clustering of galaxies in the completed {SDSS}-{III} {Baryon} {Oscillation} {Spectroscopic} {Survey}: cosmological analysis of the {DR12} galaxy sample}, \href{http://dx.doi.org/10.1093/mnras/stx721}{\emph{Monthly Notices of the Royal Astronomical Society} {\bfseries 470} (Sept., 2017) 2617--2652}.

\bibitem{eboss_collaboration_completed_2021}
eBOSS Collaboration, S.~Alam, M.~Aubert, S.~Avila, C.~Balland, J.~E. Bautista et~al., \emph{The {Completed} {SDSS}-{IV} extended {Baryon} {Oscillation} {Spectroscopic} {Survey}: {Cosmological} {Implications} from two {Decades} of {Spectroscopic} {Surveys} at the {Apache} {Point} observatory}, \href{http://dx.doi.org/10.1103/PhysRevD.103.083533}{\emph{Physical Review D} {\bfseries 103} (Apr., 2021) 083533}.

\bibitem{balkenhol_measurement_2023}
L.~Balkenhol, D.~Dutcher, A.~S. Mancini, A.~Doussot, K.~Benabed, S.~Galli et~al., \emph{A {Measurement} of the {CMB} {Temperature} {Power} {Spectrum} and {Constraints} on {Cosmology} from the {SPT}-{3G} 2018 {TT}/{TE}/{EE} {Data} {Set}}, \href{http://dx.doi.org/10.1103/PhysRevD.108.023510}{\emph{Physical Review D} {\bfseries 108} (July, 2023) 023510}.

\bibitem{aiola_atacama_2020}
S.~Aiola, E.~Calabrese, L.~Maurin, S.~Naess, B.~L. Schmitt, M.~H. Abitbol et~al., \emph{The {Atacama} {Cosmology} {Telescope}: {DR4} {Maps} and {Cosmological} {Parameters}}, \href{http://dx.doi.org/10.1088/1475-7516/2020/12/047}{\emph{Journal of Cosmology and Astroparticle Physics} {\bfseries 2020} (Dec., 2020) 047--047}.

\bibitem{choi_atacama_2020}
S.~K. Choi, M.~Hasselfield, S.-P.~P. Ho, B.~Koopman, M.~Lungu, M.~H. Abitbol et~al., \emph{The {Atacama} {Cosmology} {Telescope}: a measurement of the {Cosmic} {Microwave} {Background} power spectra at 98 and 150 {GHz}}, \href{http://dx.doi.org/10.1088/1475-7516/2020/12/045}{\emph{Journal of Cosmology and Astroparticle Physics} {\bfseries 2020} (Dec., 2020) 045--045}.

\bibitem{madhavacheril_atacama_2023}
M.~S. Madhavacheril, F.~J. Qu, B.~D. Sherwin, N.~MacCrann, Y.~Li, I.~Abril-Cabezas et~al., \emph{The {Atacama} {Cosmology} {Telescope}: {DR6} {Gravitational} {Lensing} {Map} and {Cosmological} {Parameters}},  Apr., 2023.

\bibitem{qu_atacama_2023}
F.~J. Qu, B.~D. Sherwin, M.~S. Madhavacheril, D.~Han, K.~T. Crowley, I.~Abril-Cabezas et~al., \emph{The {Atacama} {Cosmology} {Telescope}: {A} {Measurement} of the {DR6} {CMB} {Lensing} {Power} {Spectrum} and its {Implications} for {Structure} {Growth}},  Apr., 2023.

\bibitem{carron_cmb_2022}
J.~Carron, M.~Mirmelstein and A.~Lewis, \emph{{CMB} lensing from {Planck} {PR4} maps}, \href{http://dx.doi.org/10.1088/1475-7516/2022/09/039}{\emph{Journal of Cosmology and Astroparticle Physics} {\bfseries 2022} (Sept., 2022) 039}.

\bibitem{gelman_inference_1992}
A.~Gelman and D.~B. Rubin, \emph{Inference from {Iterative} {Simulation} {Using} {Multiple} {Sequences}}, \href{http://dx.doi.org/10.1214/ss/1177011136}{\emph{Statistical Science} {\bfseries 7} (Nov., 1992) }.

\bibitem{Kumar2019}
R.~Kumar, C.~Carroll, A.~Hartikainen and O.~Martin, \emph{Arviz a unified library for exploratory analysis of bayesian models in python}, \href{http://dx.doi.org/10.21105/joss.01143}{\emph{Journal of Open Source Software} {\bfseries 4} (Jan., 2019) 1143}.

\bibitem{Kavanagh2018}
B.~J. Kavanagh, D.~Gaggero and G.~Bertone, \emph{Black holes' dark dress: On the merger rate of a subdominant population of primordial black holes}, \href{http://dx.doi.org/10.1103/physrevd.98.023536}{\emph{Phys. Rev. D 98, 023536 (2018)} {\bfseries 98} (July, 2018) 023536}, [\href{https://arxiv.org/abs/1805.09034}{{\ttfamily 1805.09034}}].

\bibitem{Chen2019}
Z.-C. Chen and Q.-G. Huang, \emph{Distinguishing primordial black holes from astrophysical black holes by einstein telescope and cosmic explorer}, \href{http://dx.doi.org/10.1088/1475-7516/2020/08/039}{\emph{Journal of Cosmology and Astroparticle Physics} {\bfseries 2020} (Aug., 2019) 039--039}, [\href{https://arxiv.org/abs/1904.02396}{{\ttfamily 1904.02396}}].

\bibitem{Manshanden2018}
J.~Manshanden, D.~Gaggero, G.~Bertone, R.~M.~T. Connors and M.~Ricotti, \emph{Multi-wavelength astronomical searches for primordial black holes}, \href{http://dx.doi.org/10.1088/1475-7516/2019/06/026}{\emph{Journal of Cosmology and Astroparticle Physics} {\bfseries 2019} (June, 2018) 026--026}, [\href{https://arxiv.org/abs/1812.07967}{{\ttfamily 1812.07967}}].

\bibitem{Oguri2018}
M.~Oguri, J.~M. Diego, N.~Kaiser, P.~L. Kelly and T.~Broadhurst, \emph{Understanding caustic crossings in giant arcs: characteristic scales, event rates, and constraints on compact dark matter}, \href{http://dx.doi.org/10.1103/physrevd.97.023518}{\emph{Phys. Rev. D 97, 023518 (2018)} {\bfseries 97} (Jan., 2018) 023518}, [\href{https://arxiv.org/abs/1710.00148}{{\ttfamily 1710.00148}}].

\bibitem{Blaineau2022}
T.~Blaineau, M.~Moniez, C.~Afonso, J.~N. Albert, R.~Ansari, E.~Aubourg et~al., \emph{New limits from microlensing on galactic black holes in the mass range $10m_{\odot}<m<1000m_{\odot}$}, \href{http://dx.doi.org/10.1051/0004-6361/202243430}{\emph{A\&A 664, A106 (2022)} {\bfseries 664} (Aug., 2022) A106}, [\href{https://arxiv.org/abs/2202.13819}{{\ttfamily 2202.13819}}].

\bibitem{EstebanGutierrez2023}
A.~Esteban-Gutiérrez, E.~Mediavilla, J.~Jiménez-Vicente and J.~A. Muñoz, \emph{Constraints on the abundance of pbhs from x-ray quasar microlensing observations: Substellar to planetary mass range},  \href{https://arxiv.org/abs/2307.07473}{{\ttfamily 2307.07473}}.

\bibitem{MonroyRodriguez2014}
M.~A. Monroy-Rodríguez and C.~Allen, \emph{The end of the macho era- revisited: new limits on macho masses from halo wide binaries}, \href{http://dx.doi.org/10.1088/0004-637x/790/2/159}{\emph{The Astrophysical Journal} {\bfseries 790} (July, 2014) 159}, [\href{https://arxiv.org/abs/1406.5169}{{\ttfamily 1406.5169}}].

\bibitem{Brandt2016}
T.~D. Brandt, \emph{Constraints on macho dark matter from compact stellar systems in ultra-faint dwarf galaxies}, \href{http://dx.doi.org/10.3847/2041-8205/824/2/l31}{\emph{The Astrophysical Journal Letters} {\bfseries 824} (June, 2016) L31}, [\href{https://arxiv.org/abs/1605.03665}{{\ttfamily 1605.03665}}].

\bibitem{Lu2020}
P.~Lu, V.~Takhistov, G.~B. Gelmini, K.~Hayashi, Y.~Inoue and A.~Kusenko, \emph{Constraining primordial black holes with dwarf galaxy heating}, \href{http://dx.doi.org/10.3847/2041-8213/abdcb6}{\emph{Astrophys.J.Lett. 908 (2021) 2, L23} {\bfseries 908} (Feb., 2020) L23}, [\href{https://arxiv.org/abs/2007.02213}{{\ttfamily 2007.02213}}].

\bibitem{Kavanagh2019}
B.~J. Kavanagh, \emph{bradkav/pbhbounds: Release version},  2019.
\newblock 10.5281/ZENODO.3538999.

\bibitem{Sugimura:2020rdw}
K.~Sugimura and M.~Ricotti, \emph{{Structure and Instability of the Ionization Fronts around Moving Black Holes}}, \href{http://dx.doi.org/10.1093/mnras/staa1394}{\emph{Mon. Not. Roy. Astron. Soc.} {\bfseries 495} (2020) 2966--2978}, [\href{https://arxiv.org/abs/2003.05625}{{\ttfamily 2003.05625}}].

\bibitem{Hutsi:2019hlw}
G.~H\"utsi, M.~Raidal and H.~Veerm\"ae, \emph{{Small-scale structure of primordial black hole dark matter and its implications for accretion}}, \href{http://dx.doi.org/10.1103/PhysRevD.100.083016}{\emph{Phys. Rev. D} {\bfseries 100} (2019) 083016}, [\href{https://arxiv.org/abs/1907.06533}{{\ttfamily 1907.06533}}].

\bibitem{Inman:2019wvr}
D.~Inman and Y.~Ali-Ha\"\i{}moud, \emph{{Early structure formation in primordial black hole cosmologies}}, \href{http://dx.doi.org/10.1103/PhysRevD.100.083528}{\emph{Phys. Rev. D} {\bfseries 100} (2019) 083528}, [\href{https://arxiv.org/abs/1907.08129}{{\ttfamily 1907.08129}}].

\bibitem{Carr:2019kxo}
B.~Carr, S.~Clesse, J.~Garc\'\i{}a-Bellido and F.~K\"uhnel, \emph{{Cosmic conundra explained by thermal history and primordial black holes}}, \href{http://dx.doi.org/10.1016/j.dark.2020.100755}{\emph{Phys. Dark Univ.} {\bfseries 31} (2021) 100755}, [\href{https://arxiv.org/abs/1906.08217}{{\ttfamily 1906.08217}}].

\bibitem{Niemeyer:1997mt}
J.~C. Niemeyer and K.~Jedamzik, \emph{{Near-critical gravitational collapse and the initial mass function of primordial black holes}}, \href{http://dx.doi.org/10.1103/PhysRevLett.80.5481}{\emph{Phys. Rev. Lett.} {\bfseries 80} (1998) 5481--5484}, [\href{https://arxiv.org/abs/astro-ph/9709072}{{\ttfamily astro-ph/9709072}}].

\bibitem{Franciolini:2022tfm}
G.~Franciolini, I.~Musco, P.~Pani and A.~Urbano, \emph{{From inflation to black hole mergers and back again: Gravitational-wave data-driven constraints on inflationary scenarios with a first-principle model of primordial black holes across the QCD epoch}}, \href{http://dx.doi.org/10.1103/PhysRevD.106.123526}{\emph{Phys. Rev. D} {\bfseries 106} (2022) 123526}, [\href{https://arxiv.org/abs/2209.05959}{{\ttfamily 2209.05959}}].

\bibitem{Chisholm:2005vm}
J.~R. Chisholm, \emph{{Clustering of primordial black holes: basic results}}, \href{http://dx.doi.org/10.1103/PhysRevD.73.083504}{\emph{Phys. Rev. D} {\bfseries 73} (2006) 083504}, [\href{https://arxiv.org/abs/astro-ph/0509141}{{\ttfamily astro-ph/0509141}}].

\bibitem{Ballesteros:2018swv}
G.~Ballesteros, P.~D. Serpico and M.~Taoso, \emph{{On the merger rate of primordial black holes: effects of nearest neighbours distribution and clustering}}, \href{http://dx.doi.org/10.1088/1475-7516/2018/10/043}{\emph{JCAP} {\bfseries 10} (2018) 043}, [\href{https://arxiv.org/abs/1807.02084}{{\ttfamily 1807.02084}}].

\bibitem{Clesse:2016vqa}
S.~Clesse and J.~Garc\'\i{}a-Bellido, \emph{{The clustering of massive Primordial Black Holes as Dark Matter: measuring their mass distribution with Advanced LIGO}}, \href{http://dx.doi.org/10.1016/j.dark.2016.10.002}{\emph{Phys. Dark Univ.} {\bfseries 15} (2017) 142--147}, [\href{https://arxiv.org/abs/1603.05234}{{\ttfamily 1603.05234}}].

\bibitem{Scarcella:2023voq}
F.~Scarcella, \emph{{Black Hole Phenomenology and Dark Matter Searches}}.
\newblock Phd thesis, 11, 2023.
\newblock \href{https://arxiv.org/abs/2311.11975}{{\ttfamily 2311.11975}}.

\bibitem{Lewis:2013hha}
A.~Lewis, \emph{{Efficient sampling of fast and slow cosmological parameters}}, \href{http://dx.doi.org/10.1103/PhysRevD.87.103529}{\emph{Phys. Rev. D} {\bfseries 87} (2013) 103529}, [\href{https://arxiv.org/abs/1304.4473}{{\ttfamily 1304.4473}}].

\bibitem{Lewis:2019xzd}
A.~Lewis, \emph{{GetDist: a Python package for analysing Monte Carlo samples}},  \href{https://arxiv.org/abs/1910.13970}{{\ttfamily 1910.13970}}.

\bibitem{Neal2005}
R.~M. Neal, \emph{Taking bigger metropolis steps by dragging fast variables},  \href{https://arxiv.org/abs/math/0502099}{{\ttfamily math/0502099}}.

\bibitem{Lewis2002}
A.~Lewis and S.~Bridle, \emph{Cosmological parameters from cmb and other data: a monte-carlo approach}, \href{http://dx.doi.org/10.1103/physrevd.66.103511}{\emph{Phys.Rev.D66:103511,2002} {\bfseries 66} (Nov., 2002) 103511}, [\href{https://arxiv.org/abs/astro-ph/0205436}{{\ttfamily astro-ph/0205436}}].

\bibitem{Lewis2013}
A.~Lewis, \emph{Efficient sampling of fast and slow cosmological parameters}, \href{http://dx.doi.org/10.1103/physrevd.87.103529}{\emph{Phys. Rev. D87, 103529 (2013)} {\bfseries 87} (May, 2013) 103529}, [\href{https://arxiv.org/abs/1304.4473}{{\ttfamily 1304.4473}}].

\bibitem{Audren:2012wb}
B.~Audren, J.~Lesgourgues, K.~Benabed and S.~Prunet, \emph{{Conservative Constraints on Early Cosmology: an illustration of the Monte Python cosmological parameter inference code}}, \href{http://dx.doi.org/10.1088/1475-7516/2013/02/001}{\emph{JCAP} {\bfseries 1302} (2013) 001}, [\href{https://arxiv.org/abs/1210.7183}{{\ttfamily 1210.7183}}].

\bibitem{Brinckmann:2018cvx}
T.~Brinckmann and J.~Lesgourgues, \emph{{MontePython 3: boosted MCMC sampler and other features}},  \href{https://arxiv.org/abs/1804.07261}{{\ttfamily 1804.07261}}.

\bibitem{Mena_2019}
O.~Mena, S.~Palomares-Ruiz, P.~Villanueva-Domingo and S.~J. Witte, \emph{Constraining the primordial black hole abundance with 21-cm cosmology}, \href{http://dx.doi.org/10.1103/physrevd.100.043540}{\emph{Physical Review D} {\bfseries 100} (Aug., 2019) }.

\end{thebibliography}\endgroup

\end{document}